\documentclass[fleqn,usenatbib]{mnras}

\usepackage{newtxtext,newtxmath}

\usepackage[T1]{fontenc}

\DeclareRobustCommand{\VAN}[3]{#2}
\let\VANthebibliography\thebibliography
\def\thebibliography{\DeclareRobustCommand{\VAN}[3]{##3}\VANthebibliography}

\usepackage{graphicx}	
\usepackage{amsmath}	
\usepackage{subcaption}
\usepackage{threeparttable}
\usepackage{multirow}
\usepackage{bm}

\newcommand{\solar}{$_\odot$}
\newcommand{\zml}{$Z = 10^{-4}$}
\newcommand{\zmu}{$Z = 0.03$}
\newcommand{\solarperyr}{$_\odot \, \textrm{yr}^{-1}$}

\newcommand{\binaryc}{\texttt{binary\_c}}
\newcommand{\Mchand}{$M_{\rm Ch}$}
\newcommand{\Mwd}{$M_{\rm WD}$}

\newcommand{\Mig}{$M_{\rm ig}$}
\newcommand{\Mdot}{$\dot{M}$}

\newcommand{\Z}{$Z$}
\newcommand{\tento}[1]{$10^{#1}$}
\newcommand{\timestento}[2]{$#1 \times 10^{#2}$}

\makeatletter
\newcommand\footnoteref[1]{\protected@xdef\@thefnmark{\ref{#1}}\@footnotemark}
\makeatother

\title[Metallicity and nova populations]{The impact of metallicity on nova populations}

\author[A. Kemp et al.]{
Alex. J. Kemp$^{1,2}$\thanks{E-mail: alexander.kemp@monash.edu},
Amanda I. Karakas$^{1,2}$,
Andrew R. Casey$^{1,2}$,
Chiaki Kobayashi$^{3}$,
Robert G Izzard$^{4}$,
\\
$^{1}$School of Physics \& Astronomy, Monash University, Clayton 3800, Victoria, Australia\\
$^{2}$Centre of Excellence for Astrophysics in Three Dimensions (ASTRO-3D), Melbourne, Victoria, Australia\\
$^{3}$Centre for Astrophysics Research, Department of Physics, Astronomy and Mathematics, University of Hertfordshire, Hatfield, AL10 9AB, UK\\
$^{4}$Astrophysics Research Group, University of Surrey, Guildford, Surrey GU2 7XH, UK\\
}

\date{Accepted XXX. Received YYY; in original form ZZZ}

\pubyear{2015}

\begin{document}
\label{firstpage}
\pagerange{\pageref{firstpage}--\pageref{lastpage}}
\maketitle

\begin{abstract}

The metallicity of a star affects its evolution in a variety of ways, changing stellar radii, luminosities, lifetimes, and remnant properties. In this work, we use the population synthesis code \binaryc\ to study how metallicity affects novae in the context of binary stellar evolution. We compute a 16-point grid of metallicities ranging from \zml\ to 0.03, presenting distributions of nova white dwarf masses, accretion rates, delay-times, and initial system properties at the two extremes of our 16-point metallicity grid. We find a clear anti-correlation between metallicity and the number of novae produced, with the number of novae at \zmu\ roughly half that at \zml. The white dwarf mass distribution has a strong systematic variation with metallicity, while the shape of the accretion rate distribution is relatively insensitive. We compute a current nova rate of approximately 33 novae per year for the Milky Way, a result consistent with observational estimates relying on extra-Galactic novae but an under-prediction relative to observational estimates relying on Galactic novae. However, the shape of our predicted Galactic white dwarf mass distribution differs significantly to existing observationally derived distributions, likely due to our underlying physical assumptions. In M31, we compute a current nova rate of approximately 36 novae per year, under-predicting the most recent observational estimate of $65^{+15}_{-16}$. Finally, we conclude that when making predictions about currently observable nova rates in spiral galaxies, or stellar environments where star formation has ceased in the distant past, metallicity can likely be considered of secondary importance compared to uncertainties in binary stellar evolution.

\end{abstract}




\section{Introduction}
\label{sec:intro}

Novae are luminous transients that occur when degenerate shells of accreted material burn explosively on the surface of white dwarfs (WDs). As one of the most commonly detected optical transients, novae have the potential to be extremely useful in furthering our understanding of binary evolution \citep[e.g.,][]{starrfield2016,shara2018,darnley2019,dellavalle2020} and nucleosynthesis \citep[e.g.,][]{1981wallace,gehrz1998,jose1998,izzo2015,tajitsu2015,starrfield2020,riley2021}.

Novae can occur billions of years after the birth of their host system, and so the observed nova population of multi-generational stellar environments -- such as galaxies -- should contain nova systems with a spread in metallicities. As such, it is important to consider the effect of metallicity on nova rates and evolutionary channels. However, most theoretical nova studies focus on detailed, 1-D simulations of novae produced through the accretion of solar metallicity material. Such simulations are necessary for the calculation of important nova properties such as the critical ignition mass (\Mig, the mass of accreted material required to produce a nova eruption) and the accretion efficiency ($\eta$, the fraction of accreted material which remains on the WD after the nova eruption). These properties can then be used in population synthesis codes to study novae from a population statistics standpoint \citep[see][henceforth K21]{kemp2021}.

There are no published nova grids which adequately resolve the parameter space in terms of the white dwarf mass (\Mwd), accretion rate (\Mdot), and metallicity (\Z). \cite{chen2019} perform the first detailed evolution study directly comparing nova properties (including \Mig\ and $\eta$) at different metallicities (\zml\ and $Z=0.02$). This study is invaluable in assessing the impact that metallicity can have on detailed nova properties due to the micro-physics of the outburst. However, detailed analysis of two metallicity cases is not a substitute for a resolved metallicity grid. More importantly, there is an effect of changing the metallicity that cannot be accounted for by detailed nova evolution codes: the effect of metallicity on binary stellar evolution.

Reducing metallicity affects the evolution of a WD-forming star in a variety of ways. On the main sequence, low-mass stars experience lower core opacity as contributions from bound-free transitions reduce. To maintain hydrostatic equilibrium, the core must produce more energy compared to a higher-metallicity star, reducing main-sequence lifetimes and driving up central densities. In more massive stars, where metallicity-insensitive electron scattering dominates the core opacity, the CNO cycle dominates energy generation on the main sequence. Reducing the metallicity also lowers the abundance of CNO elements in the core, thereby decreasing the efficiency of the CNO cycle. Thus lower-metallicity stars relying on CNO cycling require higher central temperatures and densities to produce similar luminosities. Regardless of which mechanism is chiefly responsible, reducing the total metallicity results in a star that is hotter, more luminous, and smaller than a main sequence star of higher metallicity \citep{pols1998,maeder2000,heger2003,karakas2014}.

The higher main sequence central temperatures lead to more massive H-exhausted cores, impacting evolution beyond the main sequence and ultimately leading to more massive WD remnants. 
These effects can be dramatic. For example, the first (red) giant branch (FGB) may be absent for stars with \Z$\lesssim$\tento{-3} \citep[e.g.,][]{fishlock2014}, as core He burning commences on the Hertzsprung gap (HG). Higher core and shell-burning luminosities both on and beyond the main sequence typically reduce stellar lifetimes.

In low-mass binary evolution, the important point is that reducing the metallicity generally leads to shorter stellar lifetimes, smaller radii, and larger remnant masses. How these effects influence binary stellar evolution in the context of novae is discussed in Section~\ref{sec:results,metalicity}.

This work addresses the question of how intrinsic nova rates and distributions are influenced by metallicity-dependent binary stellar evolution.
We make use of population synthesis methods, deliberately neglecting the metallicity dependence of nova micro-physics \citep[e.g.,][]{kato2013,chen2019} in our model\footnote{\label{microphysics_footnote}The likely impact that accounting for metallicity-dependent micro-physics would have on our predictions is discussed in Section~\ref{sec:chen2019disc}.}. We identify how and why different evolutionary channels and distributions are affected, and link these changes to how metallicity affects stellar radii, lifetimes, and core masses. We then convolve a four-component (thin disk + thick disk + bulge + halo) model of the Milky Way (MW) based on \cite{kobayashi2020origin} and a single-component model of M31 \citep{williams2017} with a 16-point metallicity grid of population synthesis results ranging from \zml\ to $Z=0.03$ to make predictions about the modern Galactic nova population and nova rates in M31.

In Section~\ref{sec:methods} we briefly describe our population synthesis model, metallicity grid, and model galaxies. Our results relating directly to how metallicity affects nova production are presented in Section~\ref{sec:results,metalicity}. Section~\ref{sec:results,galaxies} presents our predictions for nova rates and distributions in the MW and M31. In Section~\ref{sec:discussion} we discuss our results in the context of the metallicity dependence of nova micro-physics, modern observations of nova properties, and the relative importance of metallicity on novae compared to other uncertainties in binary evolution. We summarise our findings in Section~\ref{sec:conclusion}.

\begin{table}
\begin{tabular}{ll}
Stellar Type & Description\\ \hline
LMMS & Low-mass main sequence \\
MS & Main sequence \\
HG & Hertzsprung gap \\
FGB & First giant branch \\
CHeB & Core He burning \\
EAGB & Early asymptotic giant branch \\
TPAGB & Thermally pulsing asymptotic giant branch \\
HeMS & He main sequence \\
HeHG & He Hertzsprung gap \\
HeGB & He giant branch \\
HeWD & He white dwarf \\
COWD & C/O white dwarf \\
ONeWD & O/Ne white dwarf \\
NS & Neutron star \\
BH & Black hole \\
MR & Massless remnant
\end{tabular}
\caption{Summary of the \protect\cite{hurley2000,hurley2002} evolutionary phases (stellar types) and terminology used throughout this work.}
\label{tab:evotags}
\end{table}

\section{Methods}
\label{sec:methods}
\subsection{Our metallicity grid}

Population synthesis codes, including \binaryc\ \citep{izzard2004,izzard2006,izzard2009,izzard2018binary}, are limited by the underlying sets of stellar tracks that are used either directly, through interpolation \citep{spera2015,kruckow2018,agrawal2020}, or indirectly through the use of fitting formulae \citep{zwart1996,izzard2004,belczynski2008compact,stanway2016,stevenson2017formation} to perform single-stellar evolution calculations. \binaryc\ is of BSE \citep{hurley2002} heritage, and shares its reliance upon the \cite{pols1998} stellar tracks. These tracks are computed on a grid of metallicities ranging from \zml\ to $Z=0.03$. These limits are primarily due to the numeric difficulty of evolving stellar models at extremely low metallicities, and these limits transfer directly to \binaryc.

\begin{figure}
\centering
\includegraphics[width=1\columnwidth]{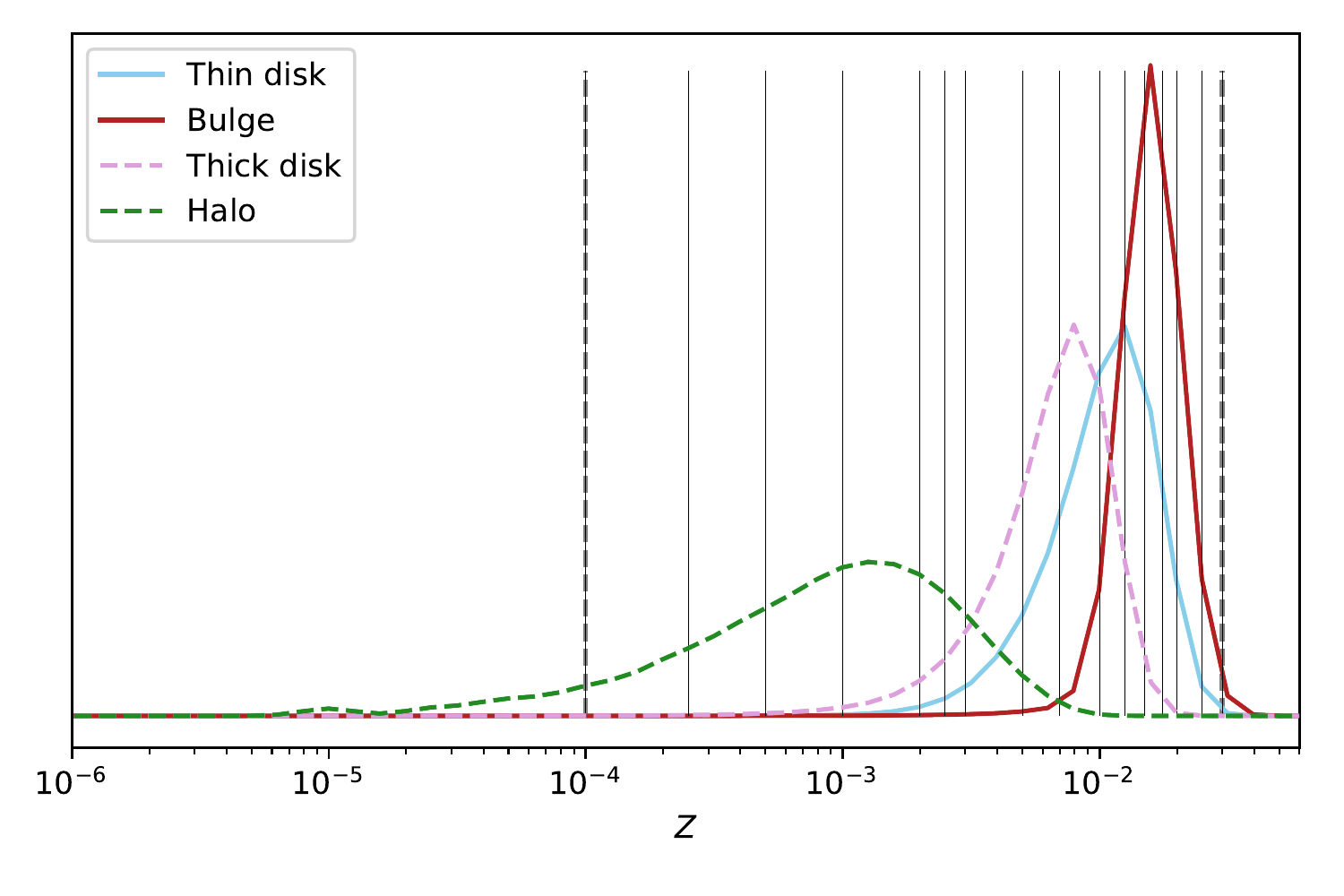}
\caption{Metallicity distribution function for the Milky Way using data from \protect \cite{kobayashi2020origin}, with our full 16-point metallicity grid overlaid. Our metallicity grid samples at \Z=\tento{-4}, \timestento{2.5}{-4}, \timestento{5}{-4},  \tento{-3}, \timestento{2}{-3}, \timestento{2.5}{-3}, \timestento{3}{-3} \timestento{5}{-3}, \timestento{7}{-3}, 0.01, 0.0125, 0.015, 0.0175, 0.02, 0.025, 0.03.}
\label{fig:mdfMW}
\end{figure}

We compute a grid of models shown by the grey lines in Figure~\ref{fig:mdfMW}, which shows the metallicity distribution function for different MW components \citep[for details, see][]{kobayashi2020origin}.  As shown in Figure~\ref{fig:mdfMW}, the (theoretical) metallicity distribution  function \citep{kobayashi2020origin} extends down to approximately \tento{-5} in the halo, and the bulge may contain stars as metal-rich as $Z=0.05$, but the vast majority of the stars reside within the metallicity bounds of \binaryc. We adopt a simple mapping scheme, where between points in our metallicity grid (and beyond its upper and lower bounds) we use data from the closest grid point, rather than interpolating or extrapolating using neighbouring points. This scheme avoids complications which arise due to the complexity of the data contained in each metallicity grid.

We emphasise that our binary physics, including our treatment of novae, remains constant between metallicity grid points. The details of our treatment of novae and our physical assumptions is described in K21 (Section 2 and Table~1) of K21.  The underlying nova models that dictate the critical ignition masses, accretion efficiencies, and accretion regimes do not change with metallicity\footnoteref{microphysics_footnote}. This is a deliberate simplification, intended to allow us to isolate the influence that the metallicity-dependence of stellar evolution plays has on novae. The numeric dimensions of each of our grids is described in Table~\ref{tab:gridH}.

This work also includes a correction to a technical error that was present in K21. In K21, the binary fraction (assumed to be 50 per cent) was incorrectly accounted for. This error manifested as a missing constant, effectively setting the binary fraction to 100 per cent. Thus only the quantitative results, namely the rate estimates for M31, were affected. These M31 rate estimates were also affected by a second error, relating to the adopted SFR history for M31 (see Section \ref{sec:methods,m31}), which acted to mitigate the impact of this error substantially.

\begin{table}
\begin{threeparttable}[b]
\begin{tabular}{lllr}
 & Bounds \tnote{a} & Spacing function & Resolution \\ \cline{2-4}
\multicolumn{1}{l|}{$M_{1,\rm init}$ (M\solar)} & (0.8, 10) & $\Delta M_{1, \rm init}=\rm const$ & 80 \\
\multicolumn{1}{l|}{$q_{\rm init}$} & ($\frac{0.1}{M_{1, \rm init}}$,1) & $\Delta (q_{\rm init})=\rm const$ & 50 \\
\multicolumn{1}{l|}{$a_{\rm init}$ (R\solar)} & (3, $1\mathrm{e}{5}$) & $\Delta\ln(a_{\rm init})=\rm const$ & 60
\end{tabular}
\begin{tablenotes}
\item[a] (min, max)
\end{tablenotes}
\end{threeparttable}
\caption{Grid bounds, spacing between grid points, and resolution ($80\times50\times60$) for each simulation.}
\label{tab:gridH}
\end{table}

\begin{figure}
\centering
\includegraphics[width=0.9\columnwidth]{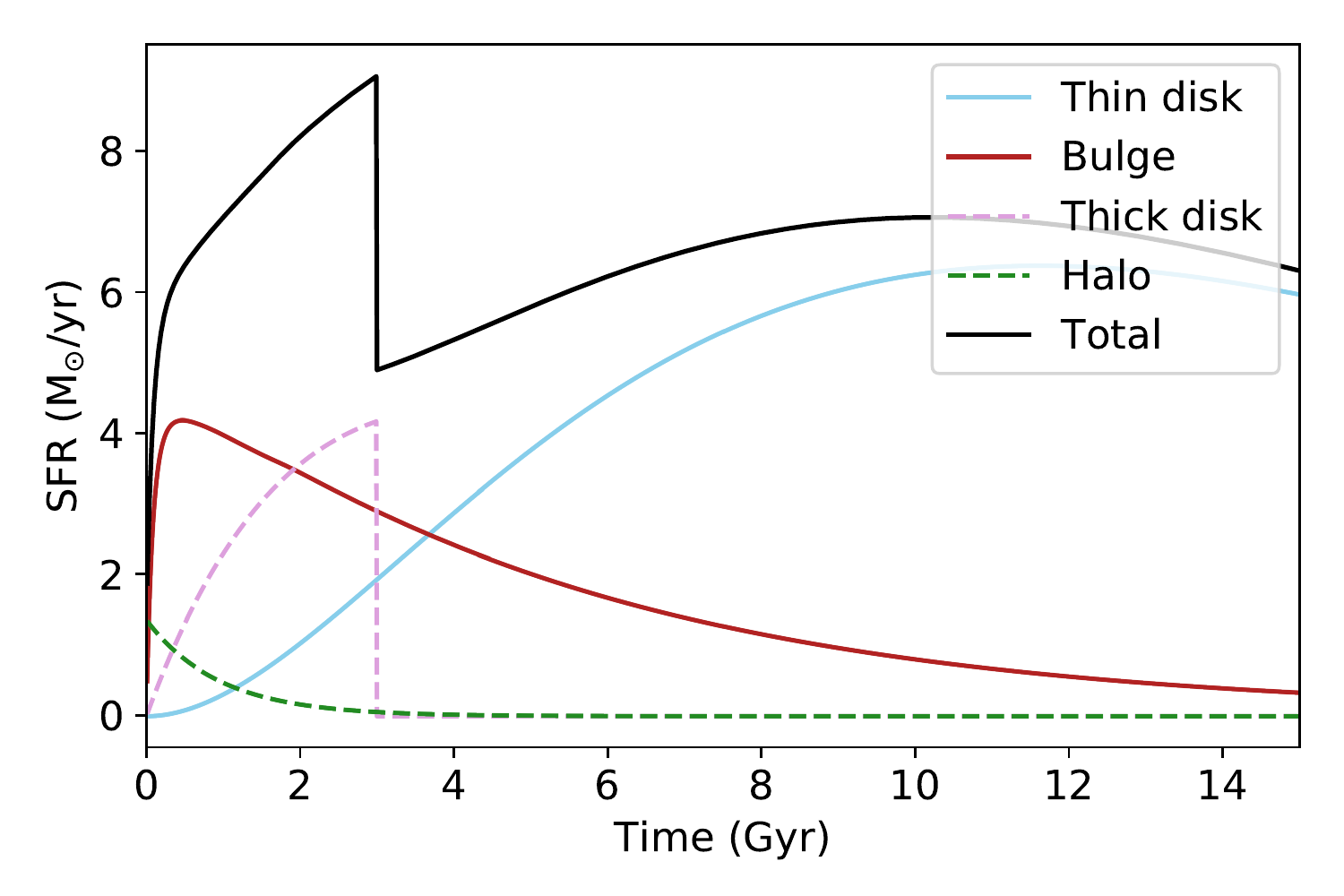}
\caption{Star formation rate history for the MW, using data from \protect \cite{kobayashi2020origin}. The total is computed as the sum of each component.}
\label{fig:sfrMW}
\end{figure}

\begin{figure}
\centering
\includegraphics[width=0.9\columnwidth]{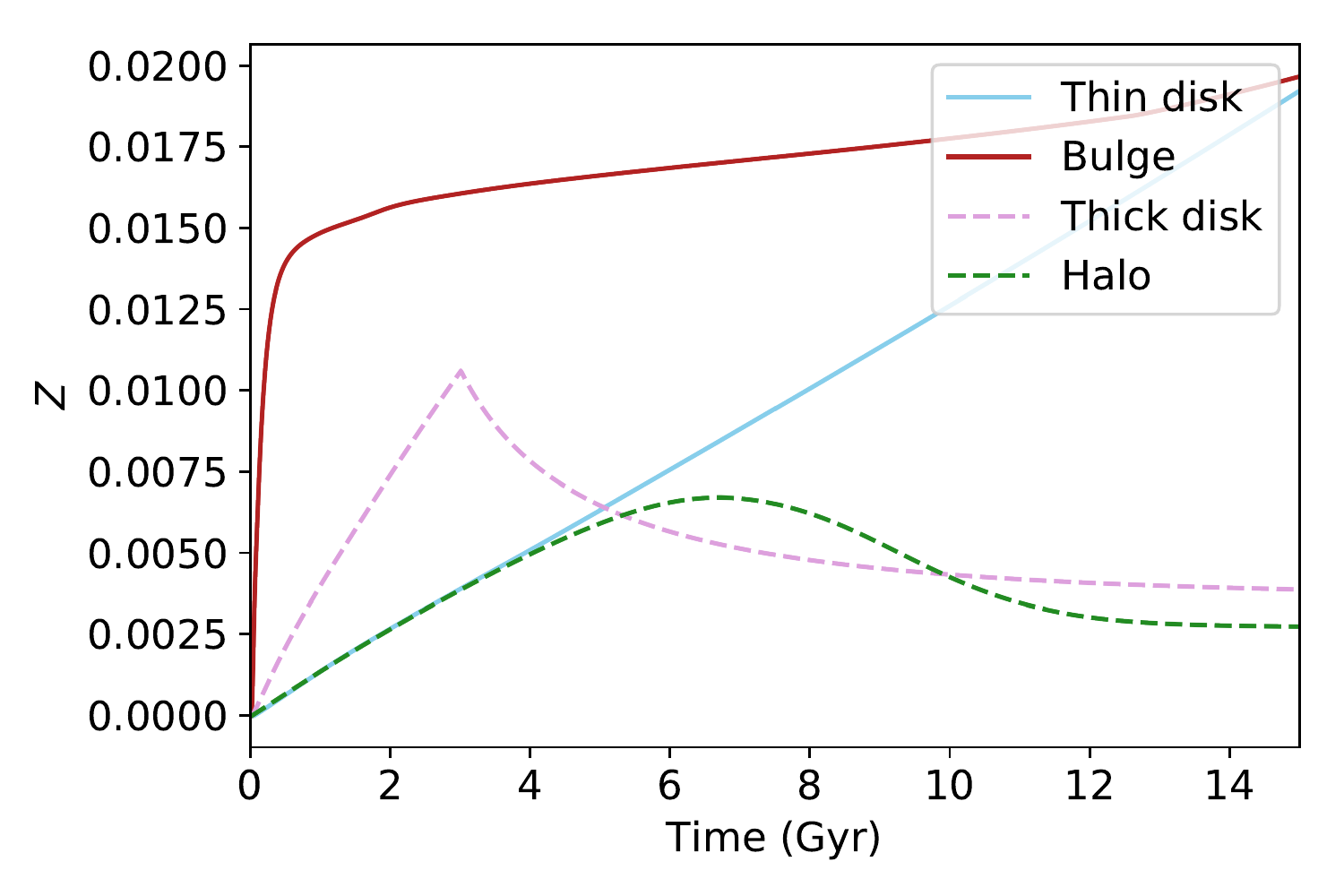}
\caption{Metallicity history for the MW, using data from \protect \cite{kobayashi2020origin}.}
\label{fig:zvtimeMW}
\end{figure}

\subsection{Our Galactic model}

To make predictions that can be compared with observations, it is necessary to build a model of the history of the desired environment. Specifically, an age-metallicity-SFR (star formation rate) relation is needed, such as that shown in Figures~\ref{fig:sfrMW} and~\ref{fig:zvtimeMW}. These relations determine, at each point in the environments lifetime, the SFR and the metallicity.

The star-forming history of any environment can be broken up into many small bursts of star formation characterised by the mass of stars formed in each starburst and the metallicity at the time of the starburst. Delay-time distributions for the desired objects of interest, computed for (approximately) the appropriate metallicity, can then be imposed originating at each starburst. This well-established method of convolving the results of population synthesis with a model galaxy was used in K21 to make predictions about the nova rate in M31, with the caveat that the metallicity component was neglected and solar-metallicity models were applied in each bin.

Here, we use a self-consistent age-metallicity-SFR relation for the Galactic thin disk, thick disk, bulge, and halo components taken from \cite{kobayashi2020origin}. These chemical evolution models well reproduce observations of the metallicity distribution function \cite[see Figure 2, panel~3 of ][]{kobayashi2020origin}, as well as other key chemical features of the MW such as [O/Fe], $\alpha$-abundance spreads, and s-process abundance trends. To calculate `true' SFRs, we assume the Galactic thin disk, thick disk, bulge, and halo to have stellar masses equal to \timestento{3.5}{10}\ M\solar, \timestento{0.5}{10}\ M\solar, \timestento{2}{10}\ M\solar, \timestento{0.8}{8}\ M\solar\ respectively \citep{blandhawthorn2016}.

The resulting SFR models are similar to the population-synthesis-friendly Galactic age-metallicity-SFR model put forth in \cite{olejak2020}'s theoretical work on galactic black hole populations. It is noteworthy that both the \cite{olejak2020} and our own Galactic model predict a current Galactic SFR of around 5-6 M\solarperyr, while observational estimates support an upper limit of not more than 2 M\solarperyr\ \citep{chomiuk2011,licquia2015,blandhawthorn2016,mor2019}. This is perhaps unsurprising, as both models are not developed to attempt to reproduce this rate, instead focusing on reproducing the observed metallicity distribution function for the Galaxy. This is the more relevant observable for the purposes of predicting current properties of events from population synthesis results, as it is star formation in the past that we are most interested in.

\subsection{Our M31 model}
\label{sec:methods,m31}
\begin{figure}
\centering
\includegraphics[width=1\columnwidth]{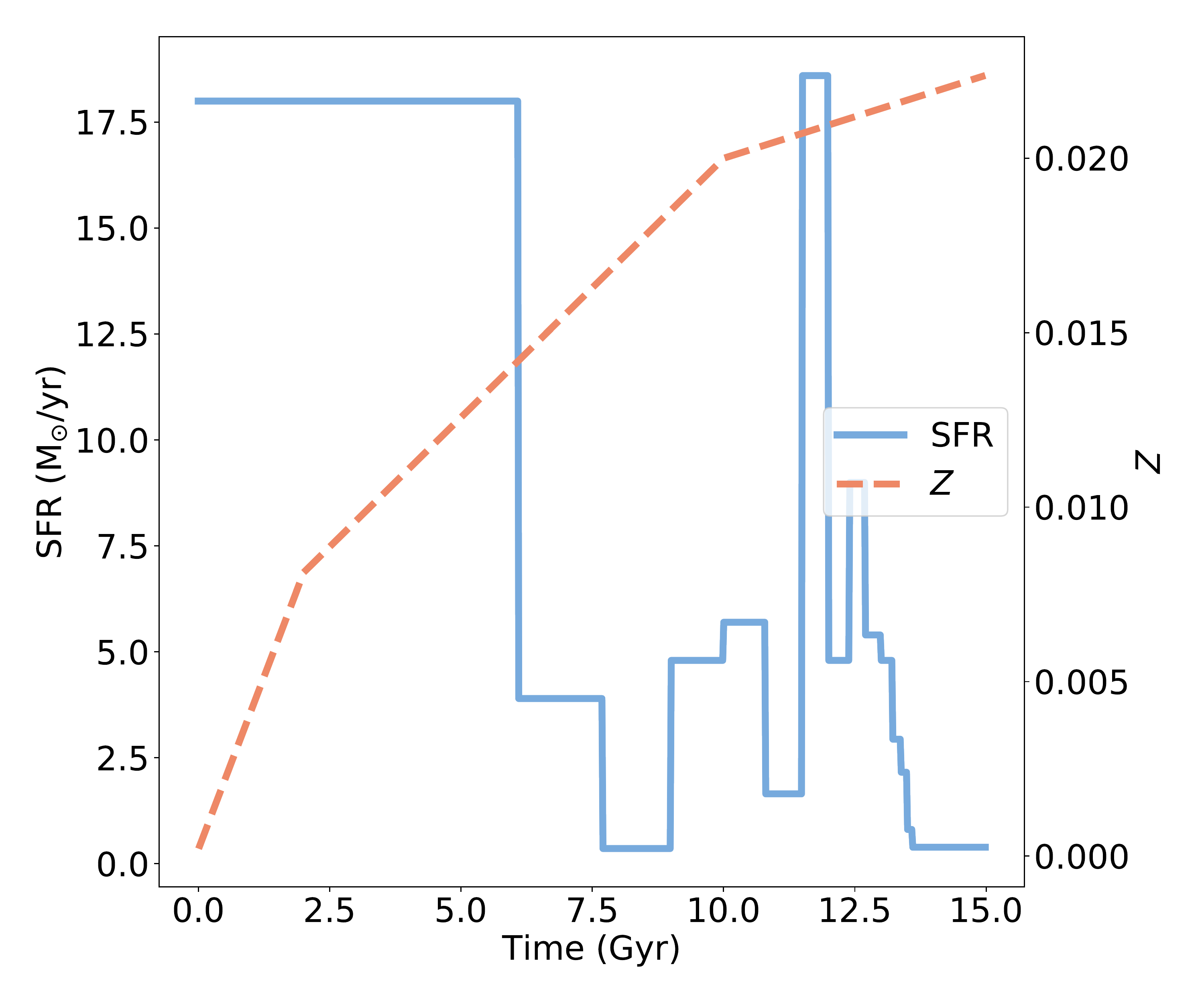}
\caption{Age-metallicity-SFR (star formation rate) for M31. The SFR is the PARSEC result from \protect \cite{williams2017}, while the metallicity history is a rough approximation to metallicity data from the same work. The implied current age of M31 is 14 Gyr.}
\label{fig:SFRvtimeM31}
\end{figure}

Our age-metallicity-SFR model for M31 is shown in Figure~\ref{fig:SFRvtimeM31}. The SFR component is taken directly from the observationally derived results of \cite{williams2017}, using their `PARSEC' result, selected based on the completeness and modernity of the underlying PARSEC stellar tracks \citep{bressan2012}. Our age-metallicity model linearly interpolates between (age (Gyr), [Fe/H]) coordinates of (0,-2), (2,-0.4), (10,0), (15,0.05) and then converts from [Fe/H] to \Z\ using $\log(Z) = 0.977 \times \textrm{[Fe/H]} - 1.699$ \citep{bertelli1994}.

Note that the use of this star formation history implies M31's current age to be 14 Gyr, as inferred by \cite{williams2017}. In K21 we assumed a 10 Gyr age for M31, curtailing the ancient star formation history by roughly 4 Gyr. Thus in that work we effectively ignored 4 Gyr of ancient star formation, resulting in an under-prediction of nova rates in M31. In this work we correct this error by taking the `PARSEC' result of \cite{williams2017} without modification, including its 14 Gyr age.

\section{The impact of metallicity on nova production}
\label{sec:results,metalicity}

In this section we discuss the impact of metallicity on nova production. However, before presenting these results it is worth considering the expected impact on nova production rates of each of the previously discussed changes brought about by varying the metallicity \textit{in isolation}. Recall that our work relies on solar-metallicity nova input models  \citep{kato2014,wang2018}; for discussion of metallicity-dependent nova micro-physics, see Section \ref{sec:chen2019disc}.

Reducing metallicity leads to shorter stellar lifetimes in low-intermediate mass stars, which has a fairly straightforward effect: stars that were insufficiently massive to evolve off the main sequence within a given time frame may now do so. This may seem a relatively unimportant change, but this has the potential to increase late-delay-time nova production significantly. K21 found that late-time nova production to be supported by a -- relatively -- large number of low-mass main sequence (LMMS) star donors producing very few novae, and a small number of FGB donors producing many novae. Thus even a small increase in the number of late-time FGB donor stars due to shorter stellar lifetimes has the potential to increase late-time nova production noticeably.

Reducing metallicity leads to increased final WD remnant masses, which also can be expected to affect nova rates. In this work we use a 3-component \cite{kroupa2001} initial mass function (IMF, see K21 for details) which we assume to be independent of metallicity. In the context of novae the important effect of this is that as metallicity reduces we produce more massive WDs at lower stellar masses. More massive WDs have lower critical ignition masses, making them more efficient producers of novae, but the effect of changing the WD mass is more complicated than that. Increasing the remnant mass also increases the influence of the WD on the binary, improving the efficiency of wind accretion and reducing the Roche radius of the companion. Due to the IMF heavily favouring the birthrates of low-mass stars, increasing the remnant masses of these objects can significantly increase nova production, particularly from lower-mass stars.

Finally, we come to perhaps the most important and least predictable of the influences of reducing the metallicity: the reduction of stellar radii. For a given WD mass, increasing the mass accretion rate leads to lower critical ignition masses, which leads to more novae. Many nova systems involve the transfer of material due to the donor star experiencing Roche-lobe overflow \cite[RLOF, see][]{paczynski1971}. The rate of this transfer is sensitive to the radial extent of the donor star, with smaller radii leading to lower mass transfer rates or potentially avoiding RLOF altogether. This would be expected to reduce nova rates. 

However, typically the reason RLOF occurs is due to the natural expansion of the donor star as it evolves. Stable RLOF -- which can be regarded as a necessary condition for RLOF-driven nova systems -- requires that the effective stellar radius attains a relatively stable equilibrium. If this condition is not met, then mass transfer is unstable and a common-envelope event will likely ensue. What this means is that, particularly for sub-giant and giant donor stars, the reduced stellar radius may not actually influence the accretion rate as greatly as one might think, as the effective stellar radius during mass transfer may actually be quite similar. The main difference in such a scenario would be that mass transfer would simply initiate further along in the evolution of the donor star.

The third influence of reducing stellar radii in terms of novae relates to common envelope physics. K21 found that only around 15\% of nova systems undergo common envelope events prior to H novae, but that these systems account for the majority of H nova events. Reducing stellar radii may lead to reducing the chances of common envelopes occurring. However, the effect of this on nova rates is unclear. The systems which undergo a common-envelope event and go on to produce novae must first have survived that common-envelope event, as systems which merge cannot produce novae. Thus there are two competing effects. Some nova systems will be `lost' -- or at least have reduced nova production -- due missing out on a common-envelope event that would have hardened the binary sufficiently for mass transfer to occur later on in the evolution of the binary. However, other systems which previously merged due to a common envelope event may instead survive to produce novae, increasing nova rates.

In summary, it is not obvious \textit{a priori} what effect changes in stellar radii due to metallicity will have on nova rates. Reducing stellar radii leads to less binary interaction, but binary evolution is complex. Avoiding or reducing the interaction at one point in time may simply delay the interaction, or allow a different interaction to take place in a later evolutionary phase when the radius is potentially orders of magnitude larger. 

\subsection{The influence of novae on nova production `per starburst'}

\label{sec:results,metalicity_sub:aggprod}

\begin{table}
\begin{tabular}{lrrr}
\textbf{\textit{Z}} & \textbf{Events / M$_{\odot \rm SFM}$} & \textbf{Systems / M$_{\odot \rm SFM}$} & \textbf{Events / System} \\ \hline
0.0001 & 11.01704 & 0.013487 & 816.8 \\
0.00025 & 9.710675 & 0.013550 & 716.6 \\
0.0005 & 9.703615 & 0.013403 & 723.9 \\
0.001 & 8.012016 & 0.012714 & 630.1 \\
0.002 & 7.732121 & 0.011855 & 652.1 \\
0.0025 & 8.298776 & 0.011382 & 729.0 \\
0.003 & 9.065271 & 0.011197 & 809.6 \\
0.005 & 5.371913 & 0.010167 & 528.3 \\
0.007 & 5.745373 & 0.009959 & 576.8 \\
0.01 & 6.836650 & 0.009579 & 713.6 \\
0.0125 & 4.359310 & 0.009069 & 480.6 \\
0.015 & 5.031115 & 0.008781 & 572.9 \\
0.0175 & 4.950579 & 0.008557 & 578.4 \\
0.02 & 5.311268 & 0.008369 & 634.5 \\
0.025 & 4.030894 & 0.007951 & 506.9 \\
0.03 & 4.594258 & 0.007649 & 600.5
\end{tabular}
\caption{Aggregate number (summed over 15 Gyr) of nova events, nova systems, and the number of nova events per system for each metallicity. The quantities are normalised per unit mass of star forming material (M$_{\odot \rm SFM}$).}
\label{tab:zsumtab}
\end{table}

\begin{figure*}
\centering

\begin{subfigure}{2.0\columnwidth}
    \centering
    \includegraphics[width=1\columnwidth]{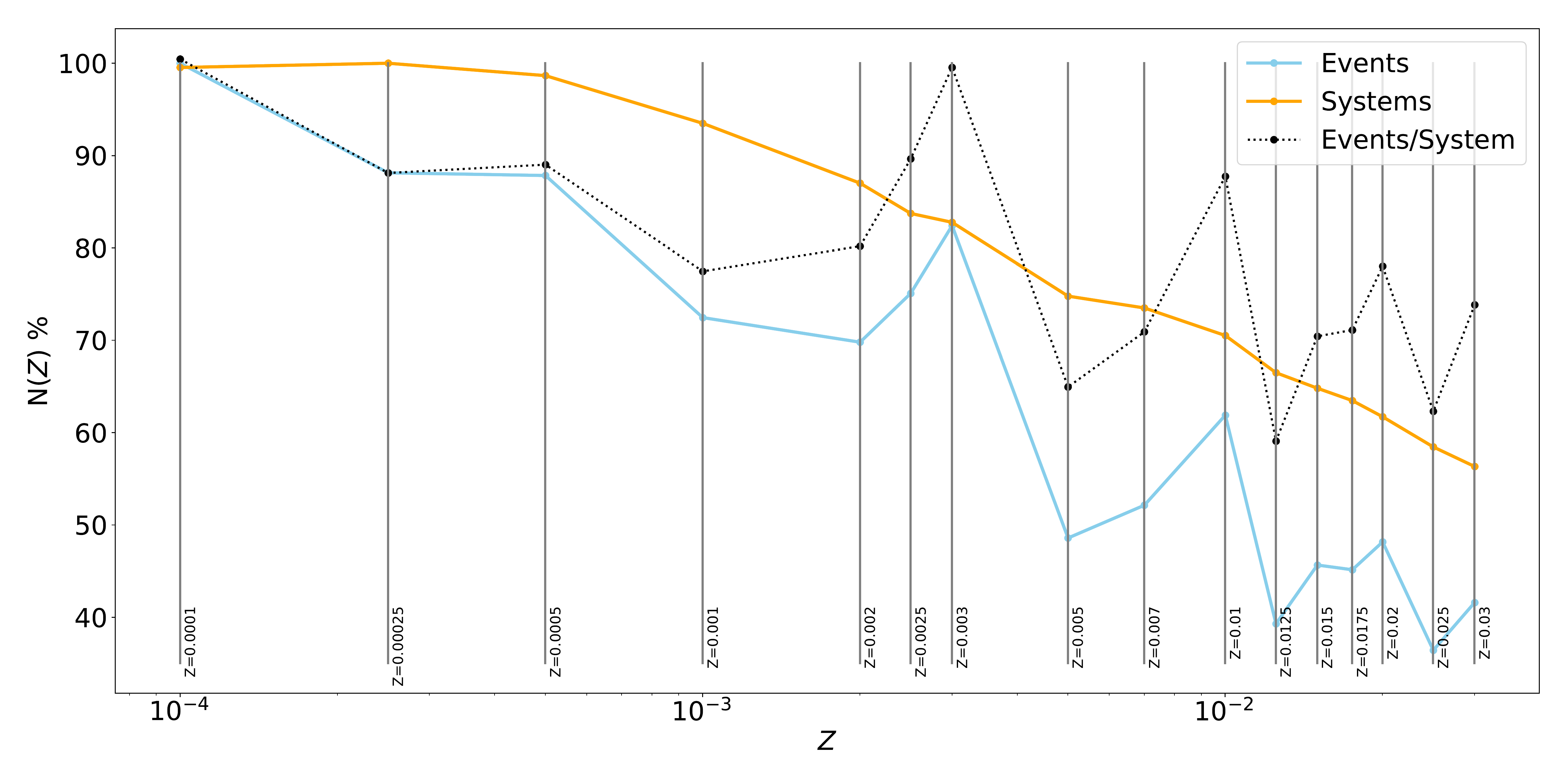}
\end{subfigure}
\caption{Fractional changes N($Z$) =$ \cfrac{X(Z)}{\mathrm{max}(X(Z))}$ of the aggregate number (summed over 15 Gyr) of nova events, nova systems, and the number of nova events per system as a function of the metallicity. There is a significant anti-correlation between all three of these quantities, with the total nova productivity from the highest metallicity systems roughly half that of the lowest.}
\label{fig:zsumtab}
\end{figure*}

Table \ref{tab:zsumtab} and Figure~\ref{fig:zsumtab} present the total number of nova events and nova systems produced during the 15 Gyr lifetime of our simulations for 1 M\solar\ simple stellar populations with fixed metallicity. As an example of what these units mean, imagine that at a given metallicity the total number of events per unit star forming mass is $A$. Were we to spend 15 Gyr perfectly observing a population of stars which formed in a single burst of star formation at this metallicity, we would detect $A\times B$ nova eruptions, where $B$ is the stellar mass of the star burst at birth.

Although this sort of information is of limited use in directly predicting the properties of most real nova populations for a multitude of reasons, it is extremely useful when considering the fundamental impacts of metallicity on nova production. Figure~\ref{fig:zsumtab} shows a clear anti-correlation between the number of nova systems formed and its metallicity. The total number of events and its ratio with the number of H-nova systems also reduces with increasing metallicity, but exhibits far more scatter.

Before addressing the trends present in Figure~\ref{fig:zsumtab}, a comment is required on the degree and source of most of the scatter present in the `Events' and `Events/System' data. We find that most of this scatter can be explained by a single, well defined evolutionary channel. In this channel, a 7-8 M\solar\ (depending on the metallicity) initial mass primary evolves without significant binary interaction, losing its envelope on the TPAGB and transferring roughly a solar mass of material onto its 1-5 M\solar\ companion, leaving a very massive ($>$1.3M\solar) WD remnant. When the companion evolves through its own TPAGB, it loses its envelope through high wind mass loss rates that lead to rapid ({\raise.17ex\hbox{$\scriptstyle\mathtt{\sim}$}}\tento{-7}\ M\solarperyr) accretion rates on the WD. This combination of high WD mass and high accretion rates leads to high nova productivity.

This channel is diverse in its spread of initial secondary masses and orbital separations, but is extremely constrained in its primary mass. The critical point is that the mass of the O/Ne WD that is formed is quite sensitive to the initial mass of the primary, and that the WD is close to the maximum WD mass able to be formed by these stars.  The critical ignition mass for these massive WDs reduces rapidly as \Mwd\ approaches \Mchand, resulting in nova productivities that can be very sensitive to \Mwd. In this particular channel, changing the primary initial mass by as little as 0.1 M\solar\ can result in up to an order of magnitude difference in the productivity of a system due to changes in the mass of the WD remnant formed.

It is this sensitivity to $M_{1,\rm init}$ that leads to most of the scatter in the `Events' data in Figure~\ref{fig:zsumtab}. As the metallicity changes, the initial mass threshold beyond which the primary is unable to form a WD remnant shifts to lower masses, which also changes the maximum mass of the WD remnants formed and, crucially, the corresponding initial mass from which they form. As in K21, we compute our results for each metallicity using a fixed grid setting out the initial primary mass, ($M_{1,\rm init}$) secondary mass ($M_{2,\rm init}$), and initial separation ($a_{\rm init}$). As we reduce the metallicity, at certain grid points we may see an over-estimated contribution from this channel if the metallicity results in the initial mass producing the maximum WD mass near-perfectly on a grid point. Conversely, if the threshold mass falls just below a grid point we may see an under-estimated contribution from this channel.

We address this through improving our effective grid resolution compared to K21 (see Table~\ref{tab:gridH}), and by carefully inspecting each of the simulations to ensure that we are not either egregiously over- or under-estimating the contribution of this channel. The results of this curating are presented in Figure~\ref{fig:zsumtab}.

Returning to the prevailing trends in Figure~\ref{fig:zsumtab}, we observe that both the number of H nova systems and the average number of events per system reduce significantly as metallicity increases, resulting in less novae. The effect of metallicity on the overall nova productivity may be considered to be a function of the number of nova systems formed per unit mass of star forming material and the average number of eruptions produced per system, with these two aspects being independent by definition.

\begin{figure*}
\centering
\includegraphics[width=2.0\columnwidth]{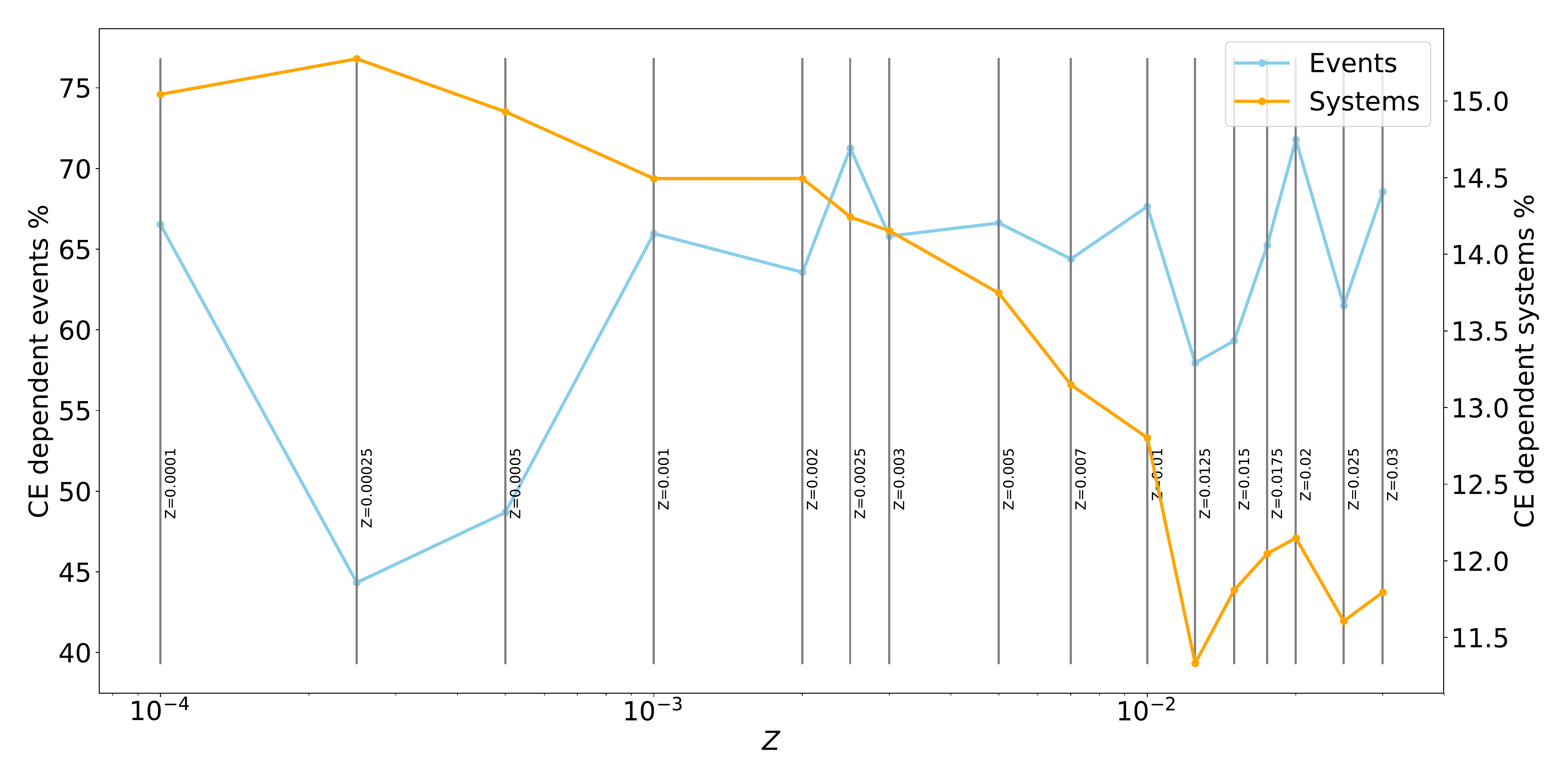}
\caption{Fraction of nova events and systems which experience at least one common envelope (CE) event occur prior to the first nova eruption. There is a strong negative trend between the fraction of CE dependent systems and the metallicity, but there are too few of these CE dependent systems for CE physics to be the main cause behind the trends seen in Figure~\ref{fig:zsumtab}.}
\label{fig:zsumtabCE}
\end{figure*}

As metallicity reduces, the number of nova systems produced per unit star-forming mass increases significantly. Common envelope physics provides one potential explanation for this, as it is possible, due to the reduction in stellar radius that occurs at lower metallicities, that CE events may be avoided or rendered more survivable, leading to binaries which would have merged prior to nova production at higher metallicities surviving to produce novae. Figure~\ref{fig:zsumtabCE} presents the fraction of nova systems (secondary $y$-axis) for which at least one CE-event occurs prior to the first nova eruption. This fraction is found to increase as metallicity reduces, but remains small (11-15 percent) across our metallicity grid. This anti-correlation implies that CE physics is contributing to the increase in nova systems, with the fraction of potentially CE-sensitive systems reducing with metallicity as stellar radii increase. Another way of putting it is that reducing stellar radii leads to a net-gain in the number of nova systems. However, the fraction of the total number of systems which are potentially CE-sensitive remains small (12-15 per cent) for all metallicities. This leads us to conclude that CE physics plays only a minor role in explaining the relationship between the number of nova systems and \Z\ shown in Figure~\ref{fig:zsumtab}.

The importance of CE physics regarding the nova eruptions themselves is also shown in Figure~\ref{fig:zsumtabCE}. With the exception of $Z=0.00025$ and $Z=0.0005$, the fraction of nova events which undergo CE events prior to nova production remains relatively flat, varying between 60 and 75 percent of all nova eruptions. Given the anti-correlation between the CE-dependent nova systems seen in Figure~\ref{fig:zsumtabCE}, it might have been expected that a similar anti-correlation would be observed in the events data. Its absence implies that the new CE-dependent nova systems introduced with reducing metallicity do not produce many novae per system, leading to any trend which might have been present being lost in the scatter of the events data in Figure~\ref{fig:zsumtab}. The apparent flatness of the events curve also indicates that CE physics does not play a significant role in the relationship between nova rates and metallicity, despite most novae being produced in systems which undergo common envelope events.

\begin{figure}
\centering
\begin{subfigure}{0.8\columnwidth}
    \centering
    \includegraphics[width=\textwidth=1]{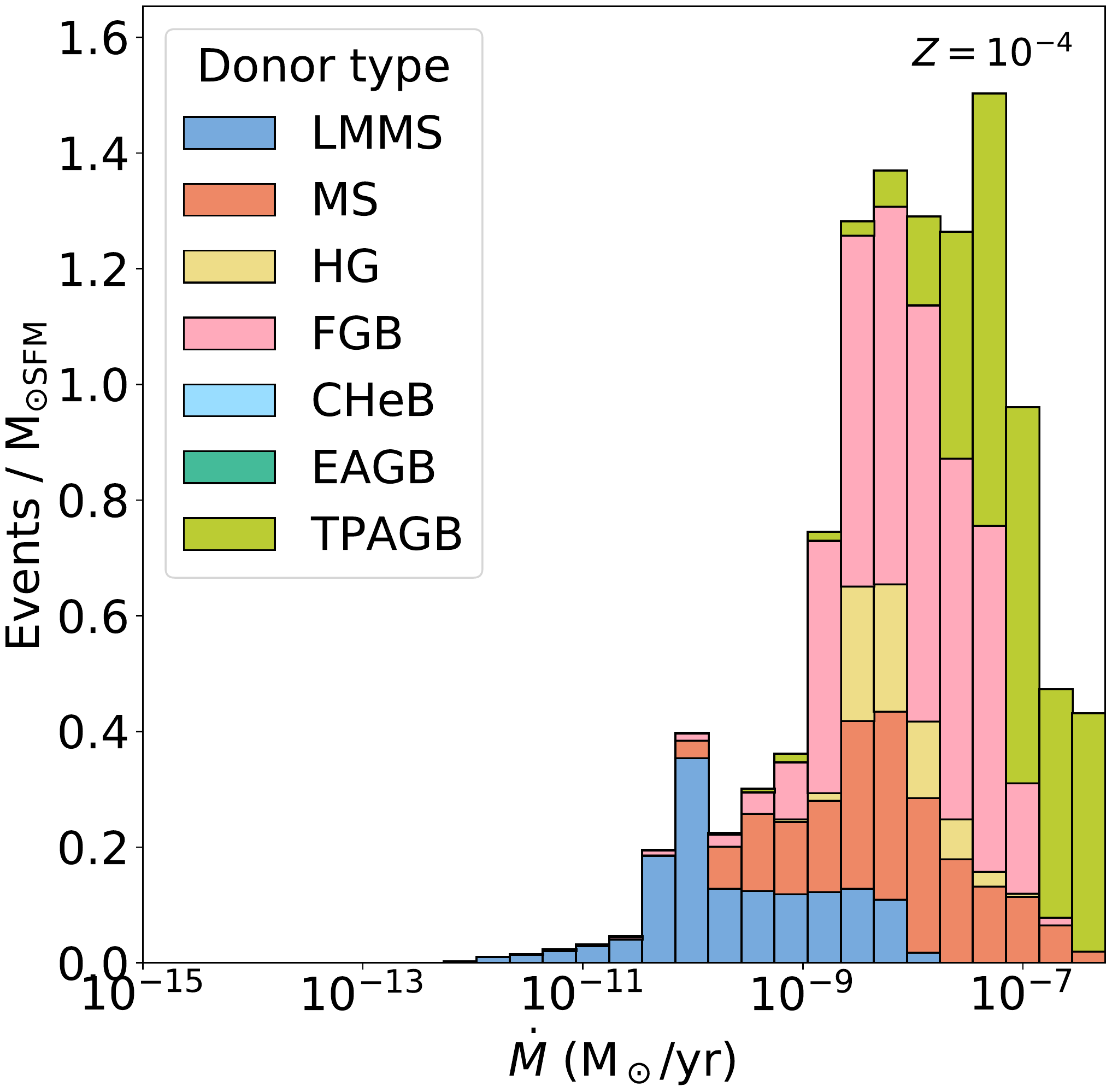}
    \caption{$Z=0.0001$}
\end{subfigure}

\begin{subfigure}{0.8\columnwidth}
    \centering
    \includegraphics[width=\textwidth=1]{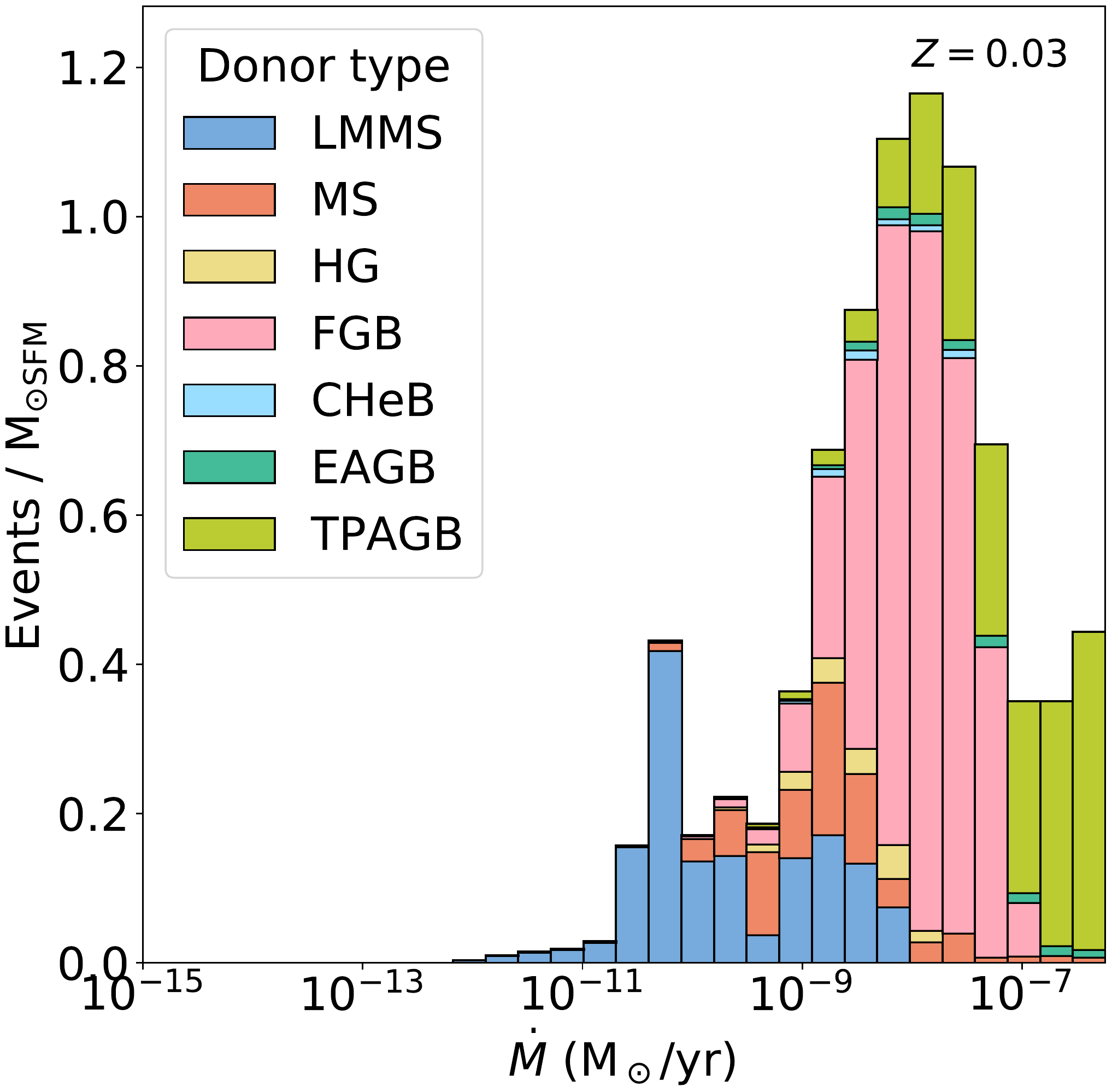}
    \caption{$Z=0.03$}
\end{subfigure}
     \centering
     \caption{Distribution of white dwarf accretion rates at the time of each nova eruption, coloured by the donor stellar type (see Table~\ref{tab:evotags} for stellar type glossary). We find no significant systematic variation in the distribution of accretion rates of novae with metallicity.}
     \label{fig:histnovamdot}
 \end{figure}

\begin{figure}
\centering
\begin{subfigure}{0.8\columnwidth}
    \centering
    \includegraphics[width=\textwidth=1]{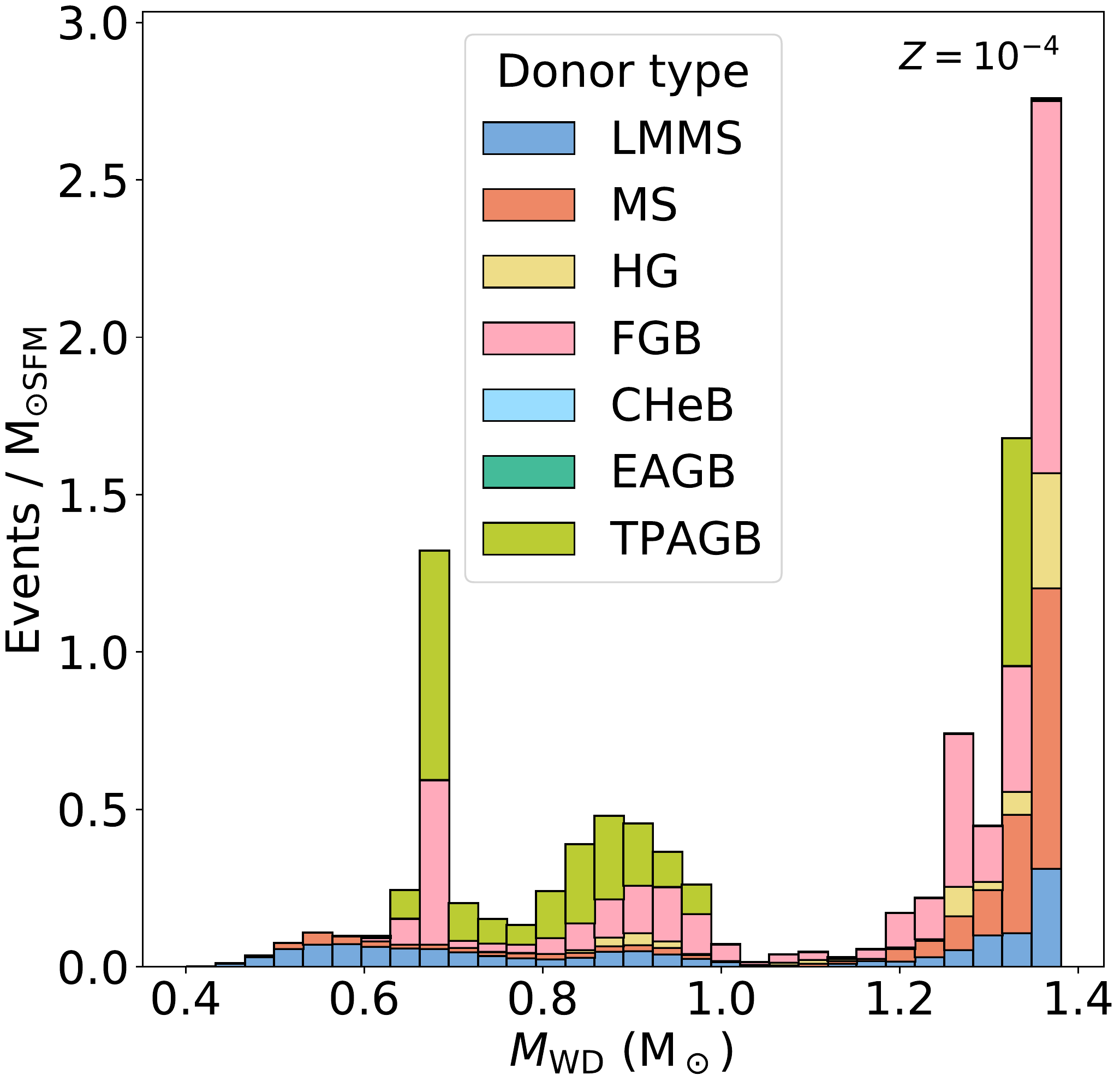}
    \caption{$Z=0.0001$}
\end{subfigure}

\begin{subfigure}{0.8\columnwidth}
    \centering
    \includegraphics[width=\textwidth=1]{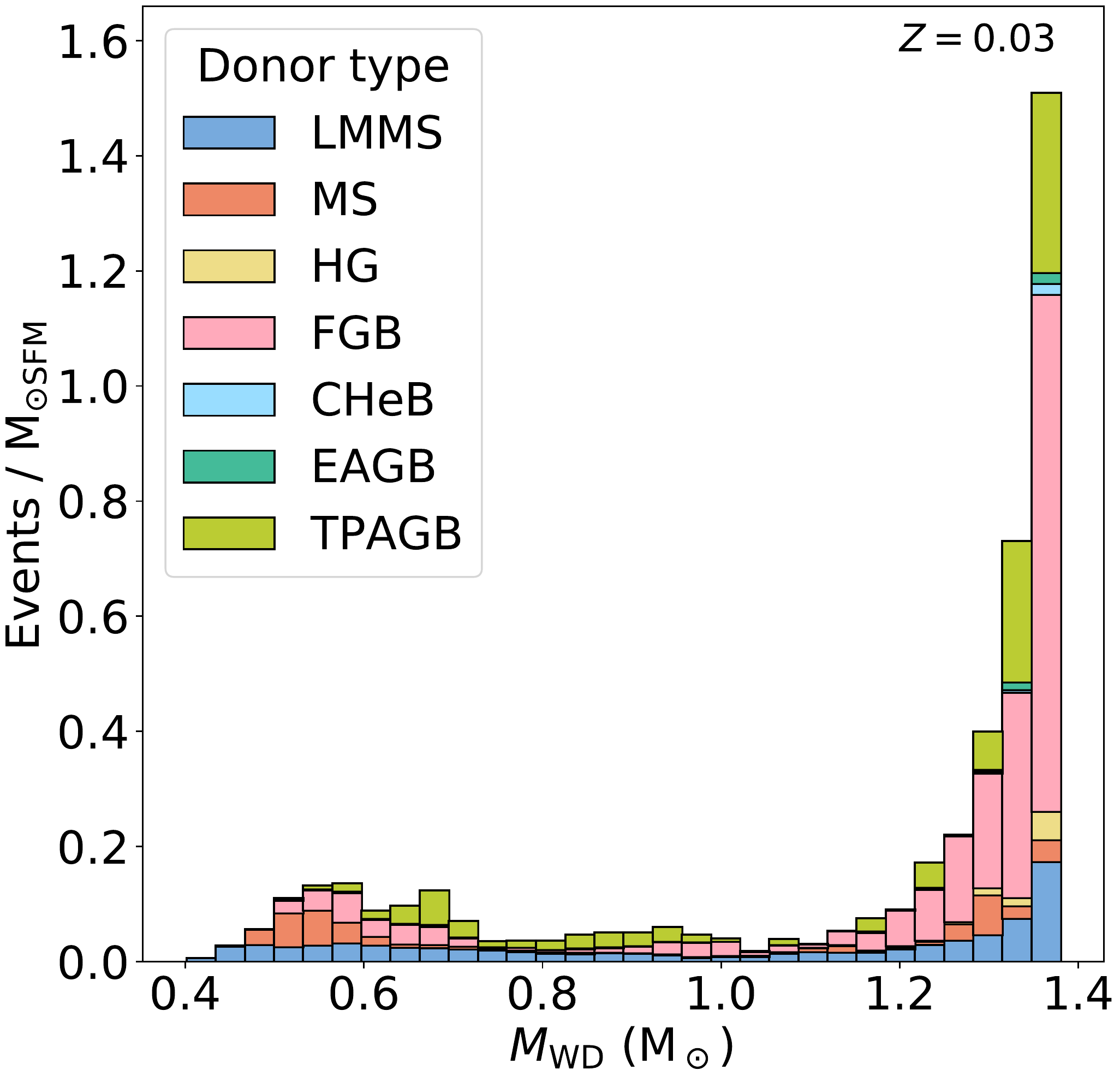}
    \caption{$Z=0.03$}
    \label{fig:histnovam1z03}
\end{subfigure}
     \centering
     \caption{Distribution of white dwarf masses at the time of each nova eruption, coloured by the donor stellar type (see Table~\ref{tab:evotags} for stellar type glossary). Particularly at lower white dwarf masses there are clear systematic differences between low and high metallicity distributions.}
     \label{fig:histnovam1}
 \end{figure}

\begin{figure*}
\begin{subfigure}{1\columnwidth}
    \centering
    \includegraphics[width=\textwidth=1]{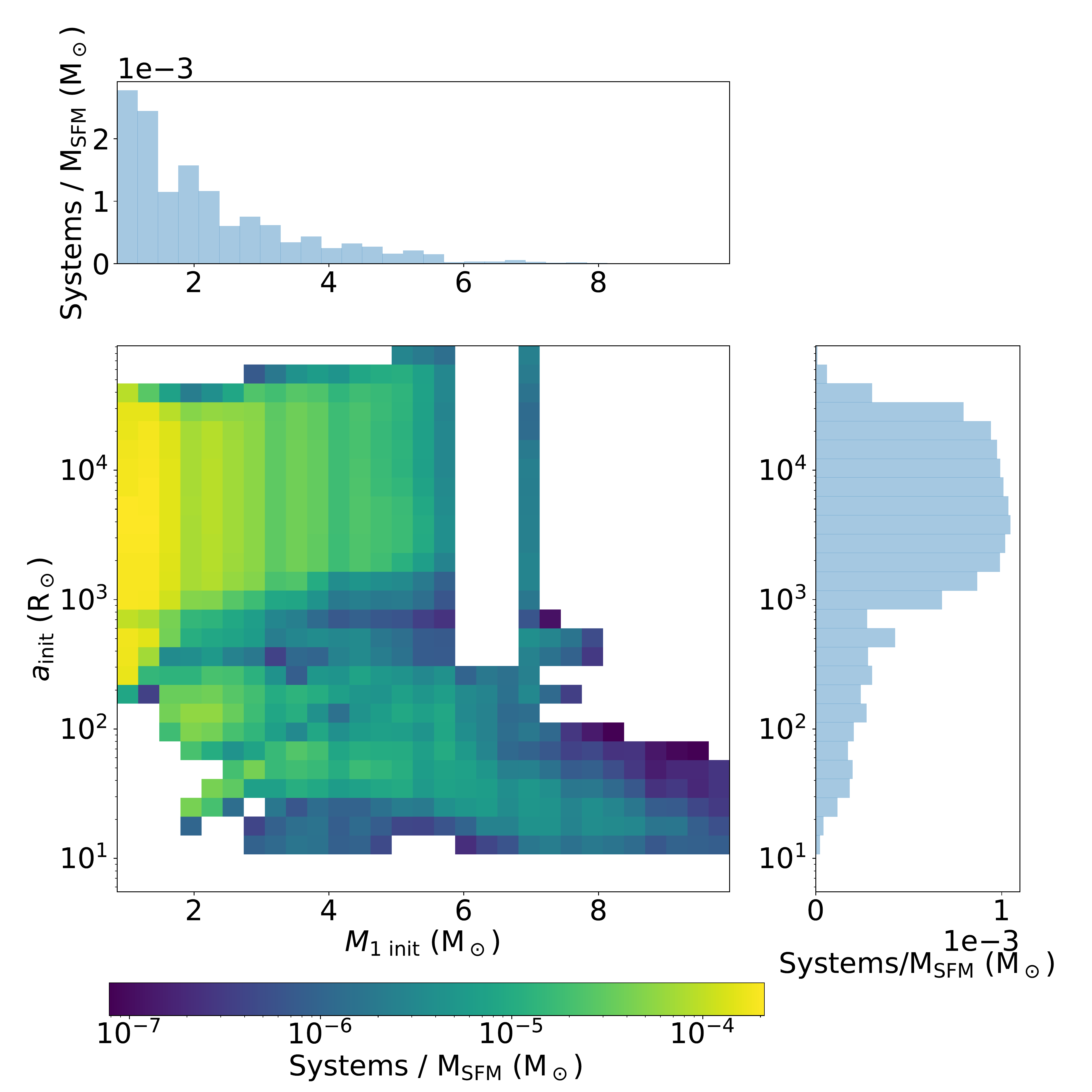}
    \caption{$Z=0.0001$}
\end{subfigure}%
\begin{subfigure}{1\columnwidth}
    \centering
    \includegraphics[width=\textwidth=1]{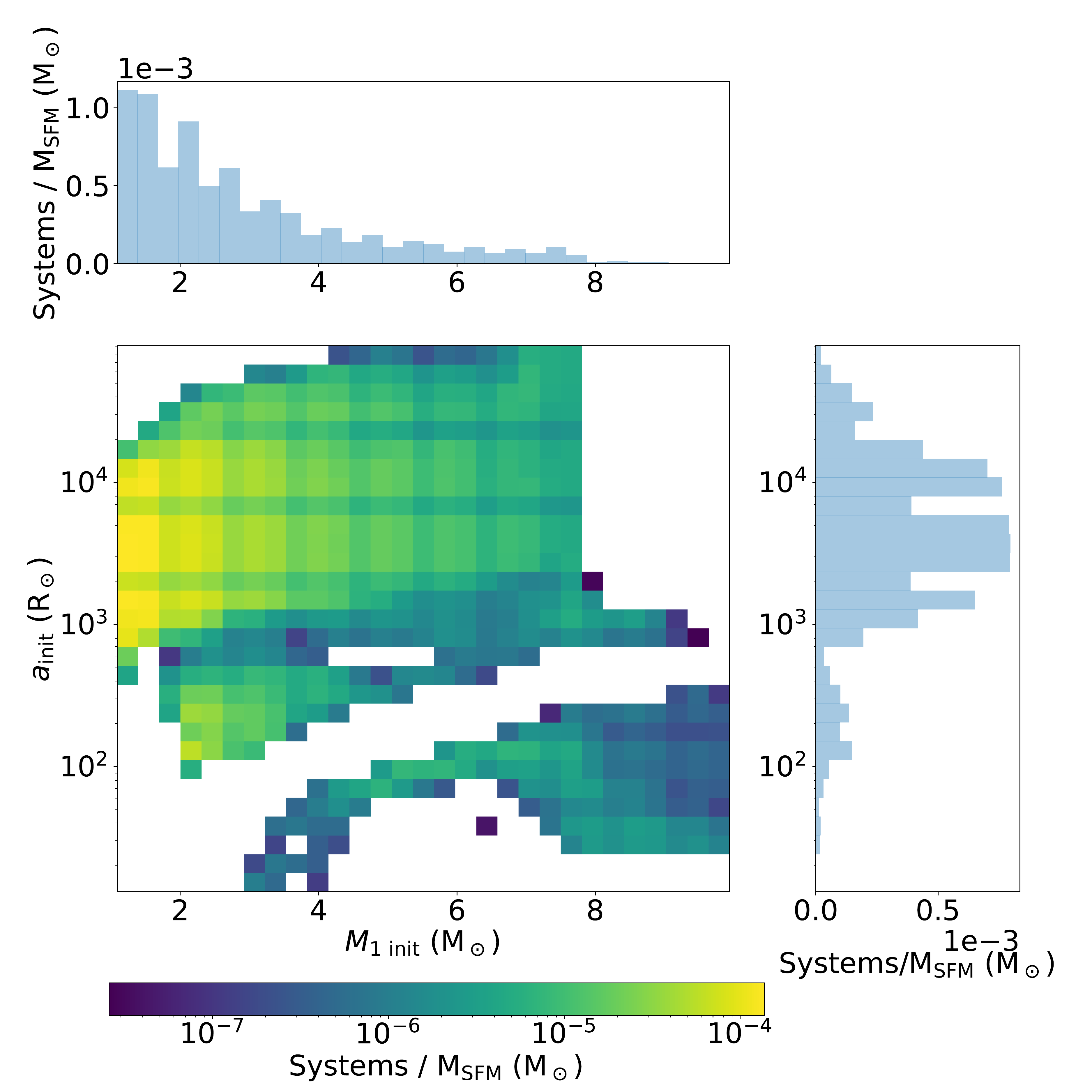}
    \caption{$Z=0.03$}
\end{subfigure}%

\caption{Distribution of initial primary mass and orbital separation, coloured by the number of nova systems produced per unit mass of star forming material in each bin. Clear variations to the morphology of the parameter space are present, discussed further in the main text.}
\label{fig:2Dhistinit_m1_a}
\end{figure*}

\begin{figure*}
\begin{subfigure}{1\columnwidth}
    \centering
    \includegraphics[width=\textwidth=1]{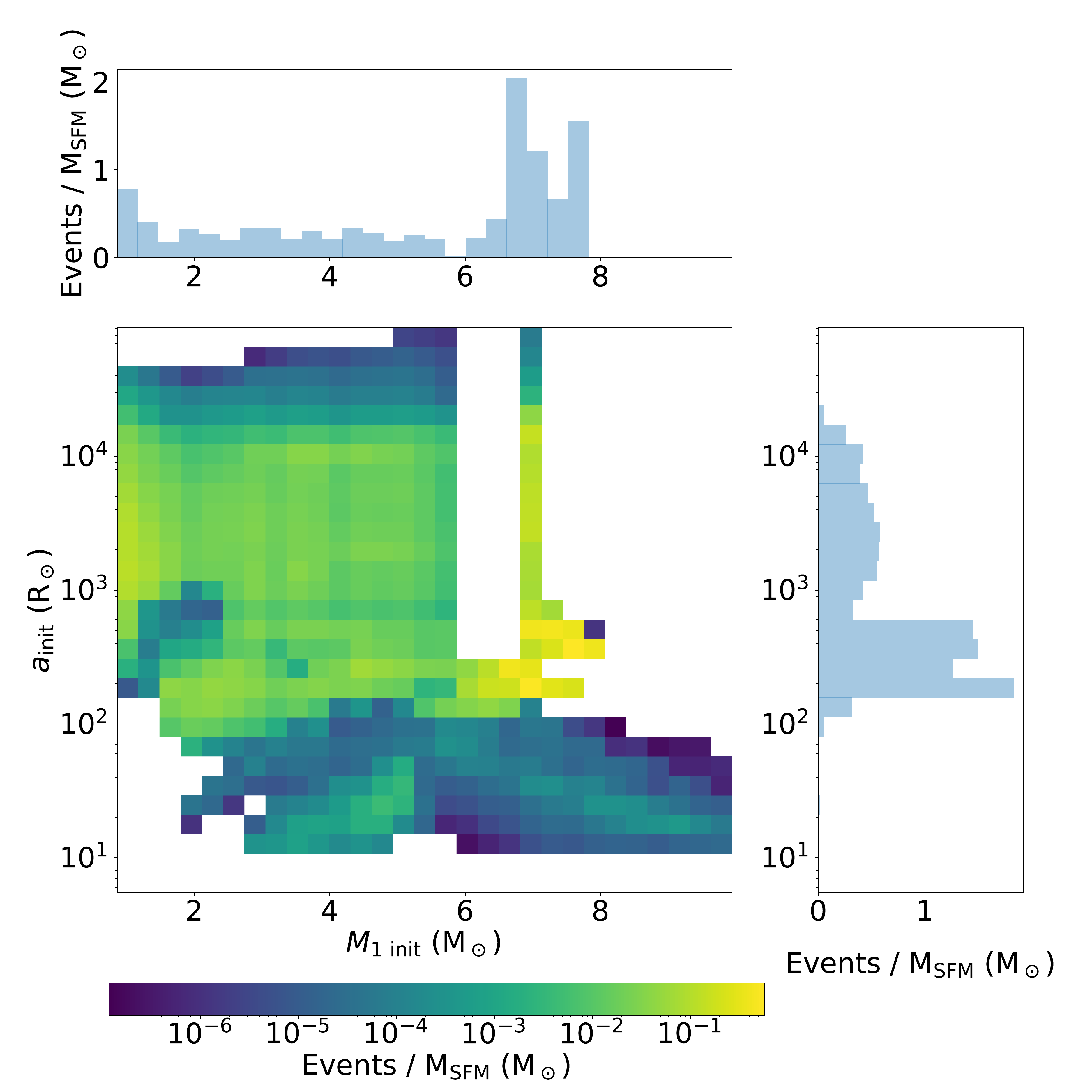}
    \caption{$Z=0.0001$}
\end{subfigure}%
\begin{subfigure}{1\columnwidth}
    \centering
    \includegraphics[width=\textwidth=1]{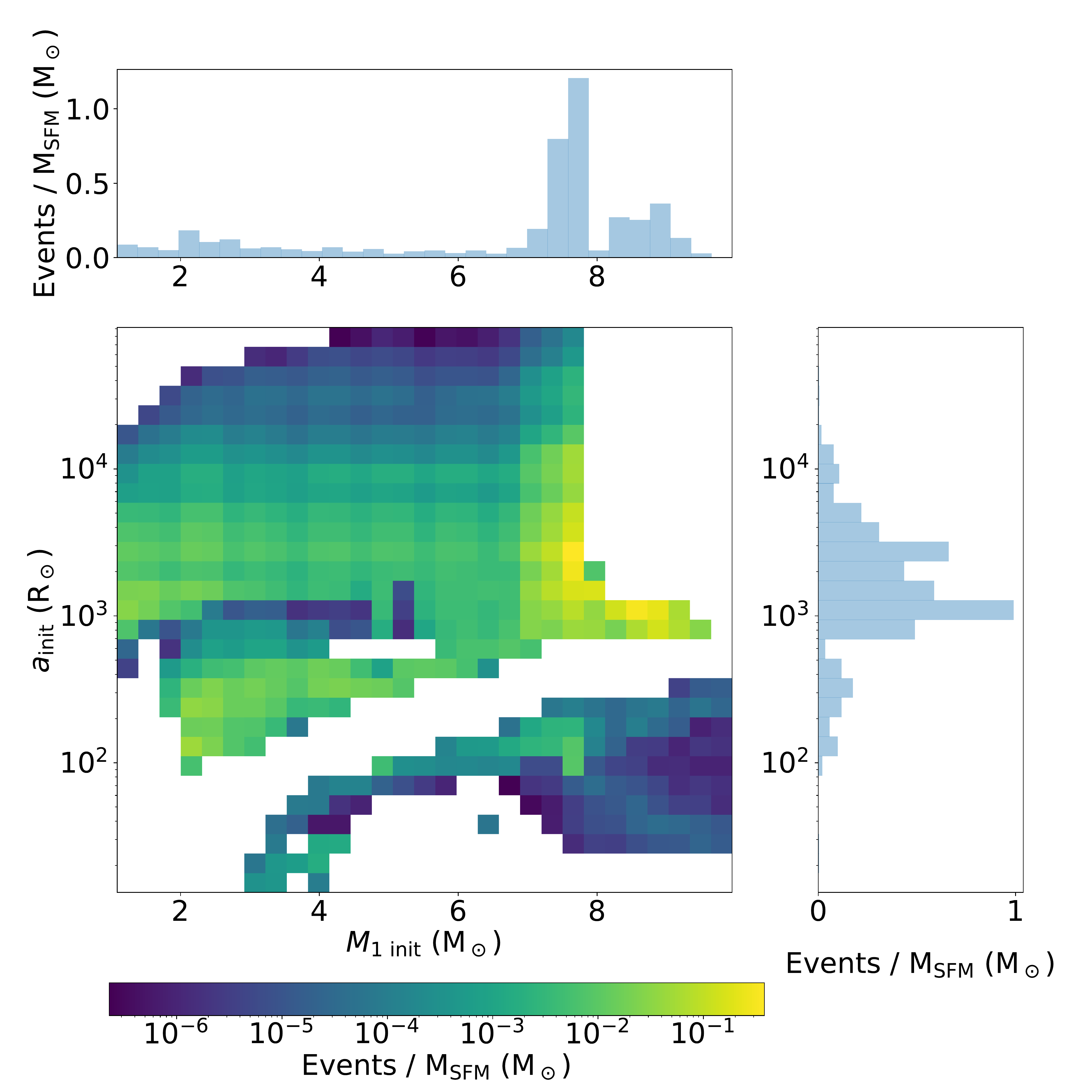}
    \caption{$Z=0.03$}
\end{subfigure}
\caption{Distribution of initial masses and orbital separations, coloured by the number of novae produced per unit mass of star forming material in each bin. Variations are present both in the distribution of where novae occur within common parameter space and the morphology of the parameter space as a whole. See main text for details.}
\label{fig:2Dhistinit_m1_aexp}
\end{figure*}

At \zml\ and \zmu\, Figures~\ref{fig:histnovamdot} and~\ref{fig:histnovam1} show the distribution of WD accretion rates and WD masses, respectively. We find no significant systematic differences between the distribution of mass accretion rates at different metallicities, superficially consistent with our previously discussed expectations regarding RLOF. We do, however, see minor systematic variation of the relative importance of certain channels. Specifically, from $Z=0.003$ and below, nova contributions from HG donors significantly increases, with a well defined, increased contribution centered -- depending on the precise metallicity examined -- around \tento{-8}\ M\solarperyr. The channel contains delay-times and component masses similar to those found for HG donor systems at higher metallicities, and relies primarily on massive WDs (\Mwd$\gtrsim 1.3$) with relatively low mass (1-2 M\solar) donor stars. Note that despite the increased prominence of HG donors at low metallicity, they remain only minor contributors compared to LMMS, MS, FGB and TPAGB donors.

Contrasting with the low amount of systematic variation in the distribution of accretion rates, Figure~\ref{fig:histnovam1} shows significant and clear variation in the distribution of \Mwd. \zml\ shows a distinctive twin peak structure, with a well defined peak around 0.65-7 M\solar\ and a second, broader peak ranging from 0.8-1.0 M\solar\, that is almost entirely absent the \zmu\ case. A complete inspection of the distributions for the entire metallicity grid reveals that contributions from WD donors between 0.6 and 1.0 M\solar\ steadily increase with decreasing metallicity between \zmu\ and \Z=\timestento{5}{-4}. This very closely mirrors the behaviour of the systems curve in Figure~\ref{fig:zsumtab}, suggesting that low-mass systems may be at the heart of this correlation.

However, Figure~\ref{fig:histnovam1} tracks nova events, not systems, and so offers direct insight into the relationship of the aggregate nova production. It is significant that the twin peaks observed in the \zml\ case are clearly driven by FGB and TPAGB donor stars. This is almost certainly a combined effect of the reduction in stellar radii and, perhaps to a lesser extent, stellar lifetimes. The principle at work here is that lowering stellar radii can avoid episodes of mass transfer or engulfment events which might prevent the WD accreting rapidly from a giant donor at higher metallicity. This could occur in a variety of ways. As metallicity reduces, channels which (at higher metallicity) produced novae with low accretion rates from main sequence donors may instead avoid significant mass transfer until later in the evolution (e.g., on the giant branch), producing significantly higher numbers of novae. A system which previously merged or was accreting too rapidly to produce novae on the giant branch may now produce significant novae. The effect of metallicity on stellar lifetimes may also contribute here, at least for very late-delay-time novae, as lower mass secondaries are now able to evolve off the main sequence, dramatically increasing the range of initial separations at which novae are possible. In short, the combined effect of reducing stellar radii and lifetimes introduces significantly more giant-donor nova systems than it removes, and it is this effect that is responsible for the twin-peak structure and likely for a significant chunk of the increased aggregate nova production at low metallicity. This increase in giant donor star contributions may be lost in the distributions of the mass accretion rates shown in Figure~\ref{fig:histnovamdot} due to the scatter from high-mass WDs affecting the distribution. 

The previously proposed relation between low-mass systems and the anti-correlation between the number of nova systems and the metallicity is supported by Figure~\ref{fig:2Dhistinit_m1_a}, which shows that at low metallicity a significant number of low-mass, longer separation systems become viable. Figure~\ref{fig:2Dhistinit_m1_a} also reveals other structural differences to the parameter space that, while not necessarily directly affecting the trends observed in Figure~\ref{fig:zsumtab}, are worthy of a very brief discussion of their origin. At \zmu, a well defined gap at initial separations around $~10^{2}$ R\solar\ is clearly visible, caused by a region of the parameter space where CE events become both inevitable and unsurvivable (see discussion on the nova `desert' in K21). This gap is steadily eroded as metallicity reduces, as CE events become more avoidable as radii reduce. This reduction in stellar radii also causes the parameter space to extend down to 10 R\solar\ at \zml, filling in and extending beyond the lower limits of the initial separations present at higher metallicities. The conspicuous gap present in the parameter space of \zml\ for initial separations $\gtrsim$\timestento{2}{2} R\solar\ and initial primary masses from 5.8-7 M\solar\ is caused by the regime where electron-capture supernovae prevent the formation of a WD remnant from the primary. This feature appears only below \Z$=$\timestento{5}{-3}, and widens with decreasing metallicity.

Returning to the low-mass, long-separation systems that appear at low metallicity, we note that their appearance may seem counter-intuitive. We have previously discussed how lowering the metallicity results in a reduction in the maximum radial extent of a given evolutionary stage. Thus it might be expected that the upper bound of $a_{\rm init}$ in Figure~\ref{fig:2Dhistinit_m1_a} should reduce with metallicity. In fact, the opposite appears to be the case. The reason for this is that in this regime, novae are almost exclusively produced through wind accretion from TPAGB donors. The significance here is that for wind accreting systems, the radius of the donor star is far less important than the relative influence of the WD accretor. Reducing the metallicity leads to more massive WD remnants, and a more massive WD companion accumulates the wind of the secondary more efficiently due to its increased gravitational influence, and thus is more likely to accumulate sufficient material to undergo at least one nova eruption. Increasing the WD mass also leads to a reduction in the critical ignition mass required for novae, so that less accreted material is required. However, the number of novae produced by these new systems is relatively small, as shown in Figure~\ref{fig:2Dhistinit_m1_aexp}.

Figure~\ref{fig:2Dhistinit_m1_aexp} also shows that the number of novae produced from WD progenitors from $1<M_{\rm 1 \ init}<6$ (corresponding to WD masses from around 0.6-1 M\solar) exhibits an increase in nova productivity over most of the parameter space. There is little difference in the distributions of secondary masses (not shown); this increase is due to a widespread, systematic increase in nova productivity on a per-system basis. The donor-driven `twin peak' low-mass WD feature (see Figure \ref{fig:histnovam1}) that appears at low metallicities is primarily responsible for this behaviour.

However, we find that this low-metallicity enhancement of giant-donor contributions at lower WD masses cannot explain all the changes in nova productivity. Nova contributions from systems with a wide range of WD masses and donor stellar types show systematic enhancements as metallicity reduces. This behaviour can be explained by the systematically higher WD remnant masses that are produced at low metallicities. Previously we discussed the introduction of low initial mass, long initial separation systems that are introduced at low metallicity due to the increased WD mass. These systems produce very few novae per system (see Figure~\ref{fig:2Dhistinit_m1_aexp}), and so their introduction acts to decrease the average number of novae per system plotted in Figure~\ref{fig:zsumtab}. 
In general, however, increasing the WD mass results in a higher number of novae per system because of the reduced critical ignition masses and the increased influence of the accretor on the binary.

We conclude that the anti-correlation between nova production and metallicity shown in Figure \ref{fig:zsumtab} is caused by two main influences. As metallicity reduces, reductions in stellar radii cause significant changes to the binary evolution of low-mass systems in particular. Further, there exists a broad increase in nova production that appears to be primarily driven by the systematically higher remnant masses at lower metallicities.

\subsection{The effect of metallicity on nova delay-time distributions}
\label{sec:results,metalicity_sub:dtd}

\begin{figure*}
\centering
\begin{subfigure}{0.9\columnwidth}
    \centering
    \includegraphics[height=0.9\textwidth]{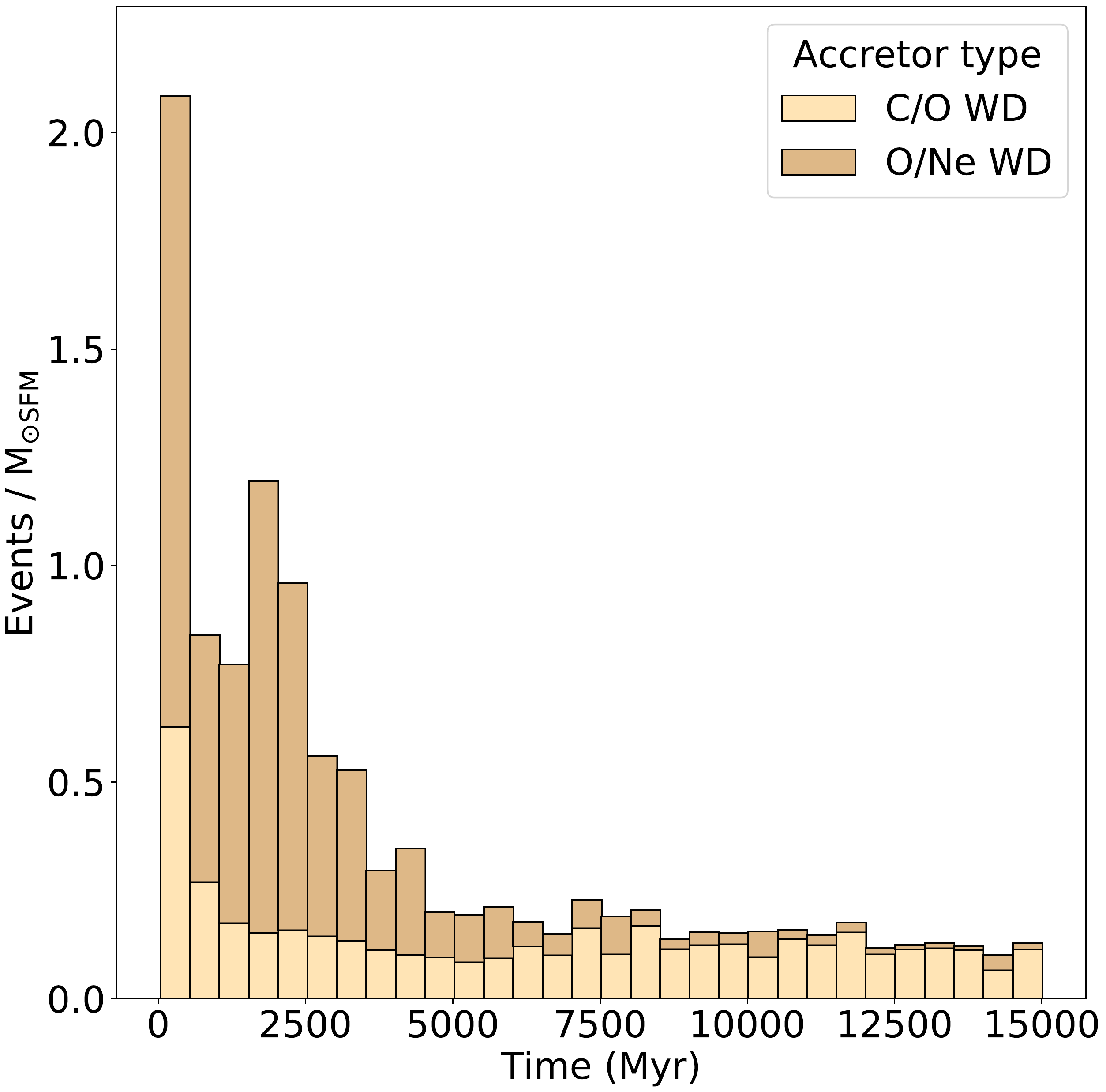}
    \caption{\Z=0.0001}
\end{subfigure}%
\begin{subfigure}{0.9\columnwidth}
    \centering
    \includegraphics[height=0.9\textwidth]{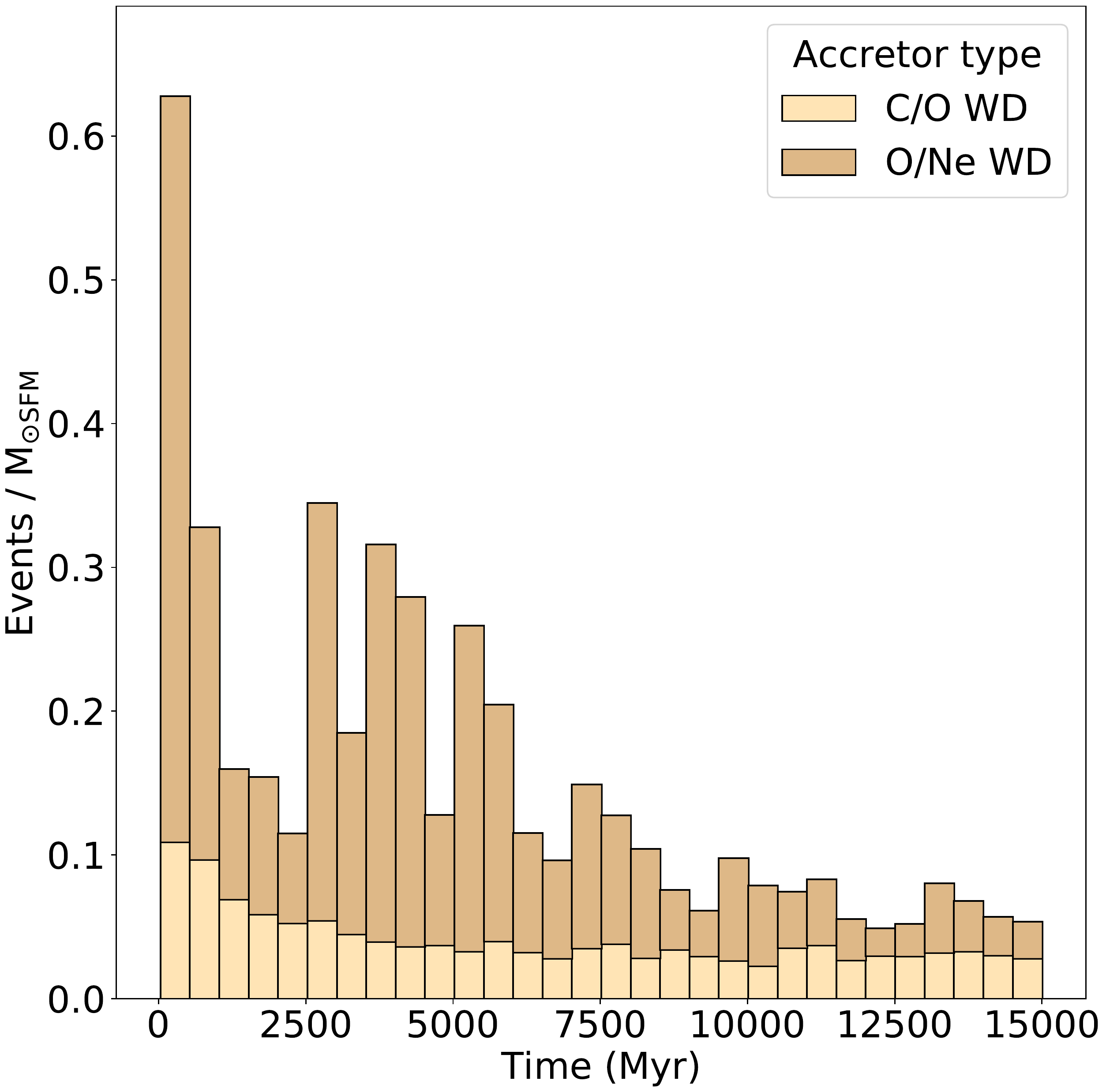}
    \caption{\Z=0.03}
\end{subfigure}

\begin{subfigure}{0.9\columnwidth}
    \centering
    \includegraphics[height=0.9\textwidth]{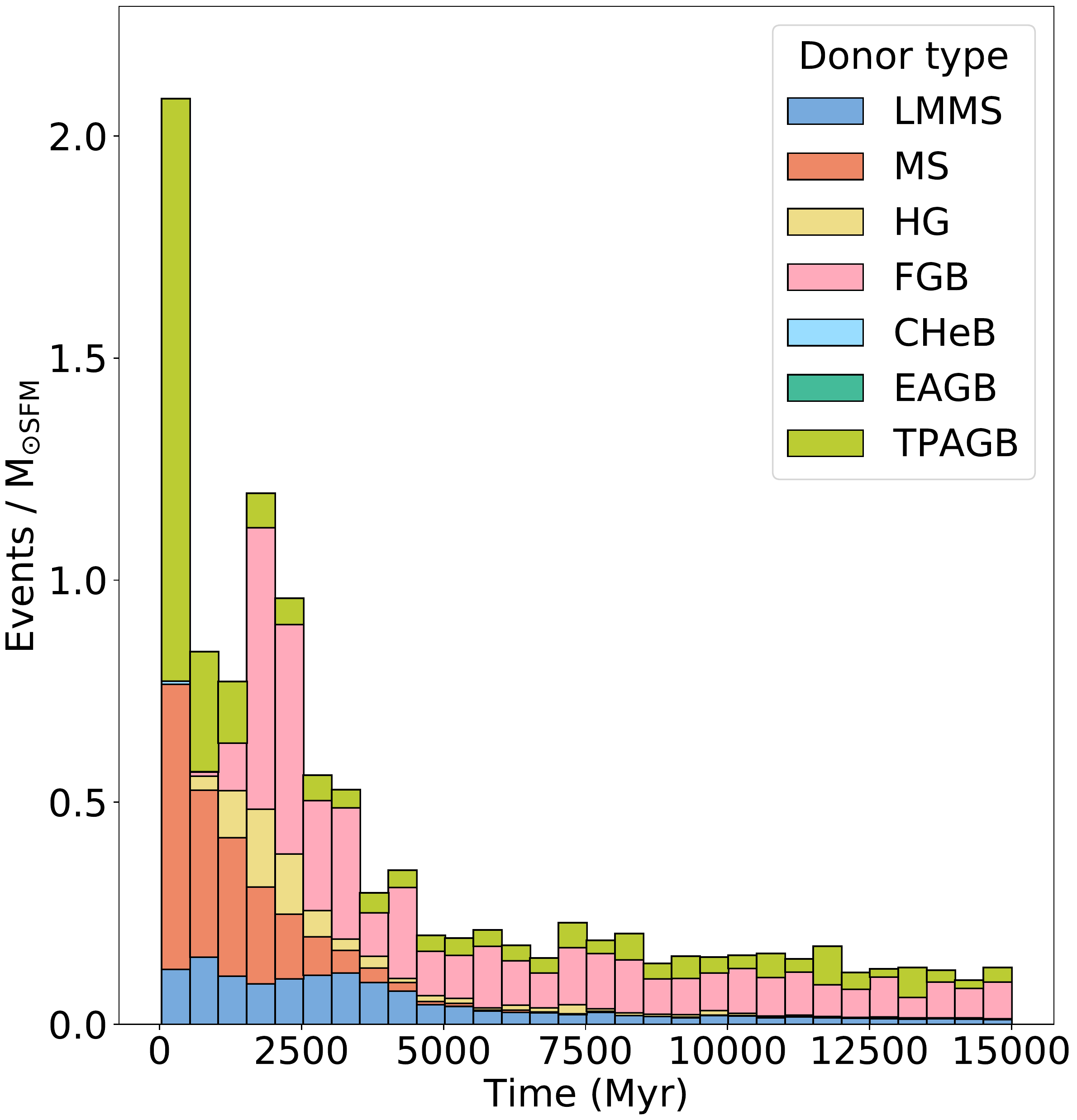}
    \caption{\Z=0.0001}
\end{subfigure}%
\begin{subfigure}{0.9\columnwidth}
    \centering
    \includegraphics[height=0.9\textwidth]{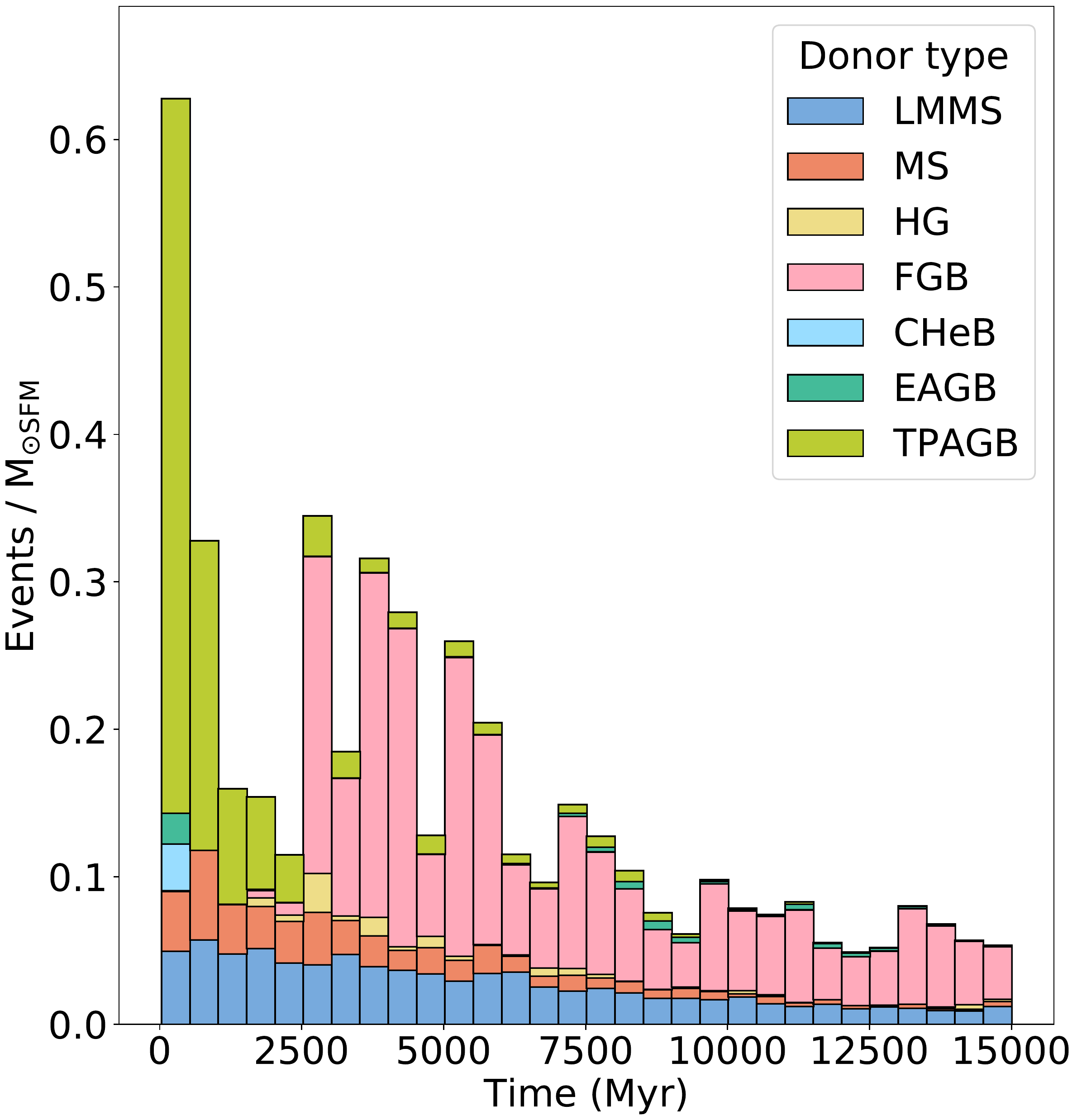}
    \caption{\Z=0.03}
\end{subfigure}

\begin{subfigure}{0.9\columnwidth}
    \centering
    \includegraphics[height=0.9\textwidth]{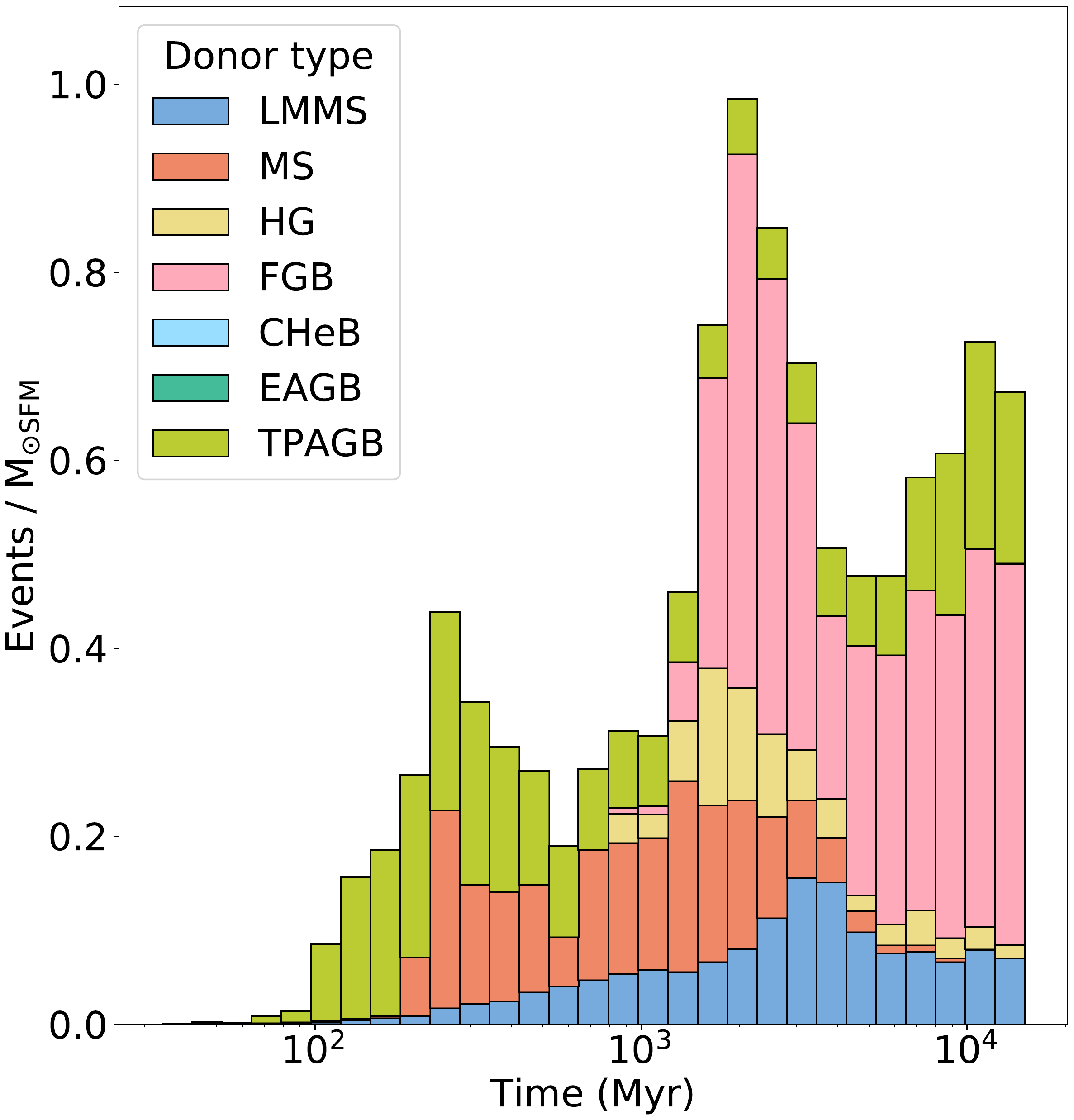}
    \caption{\Z=0.0001}
\end{subfigure}%
\begin{subfigure}{0.9\columnwidth}
    \centering
    \includegraphics[height=0.9\textwidth]{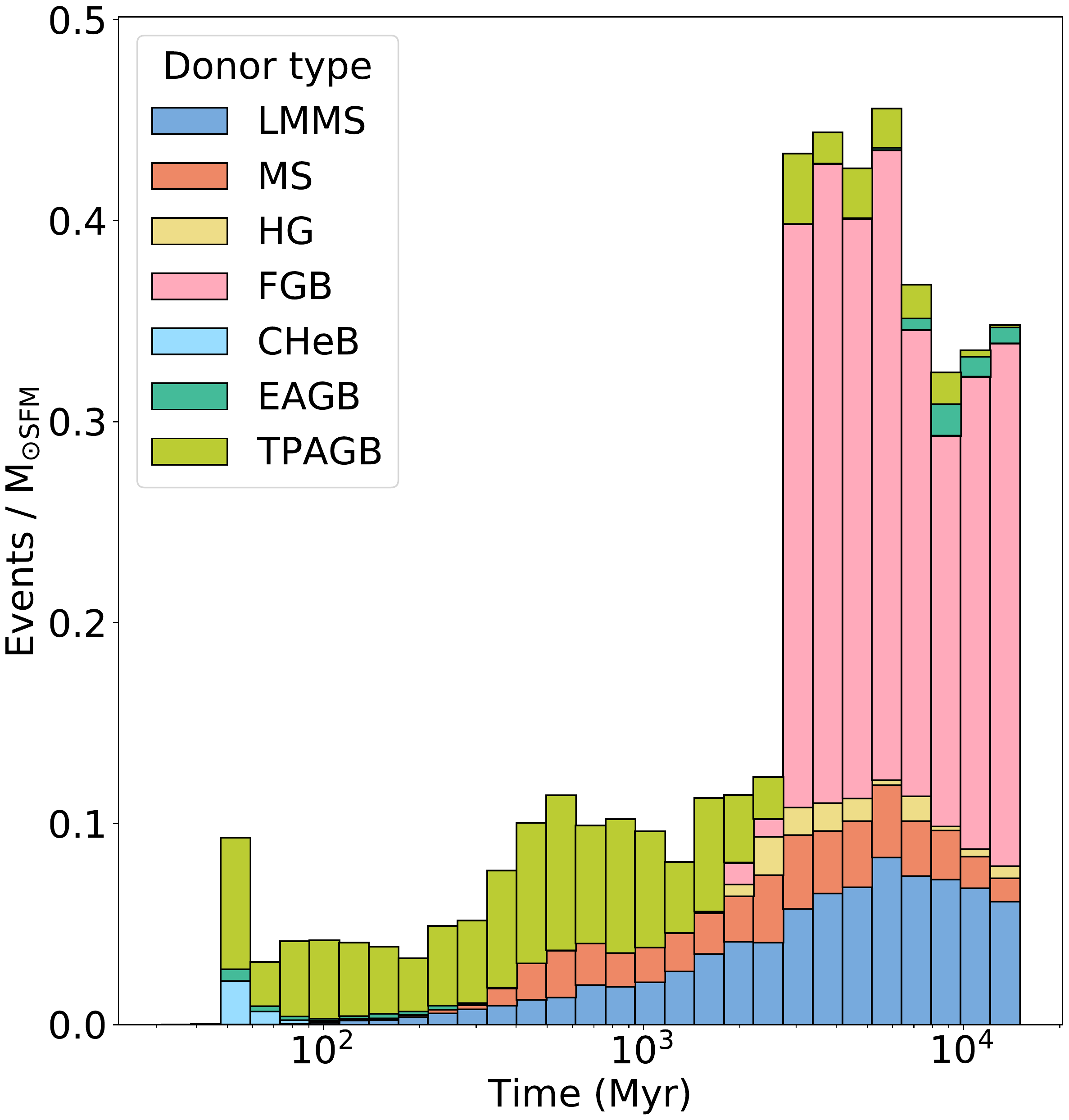}
    \caption{\Z=0.03}
\end{subfigure}
     \centering
     \caption{Distribution of nova delay-times, coloured by the donor and accretor stellar types (see Table~\ref{tab:evotags} for stellar type glossary). Star formation occurs in a single burst at time=0. Mild systematic variations in the morphology are present, driven primarily by the changes in evolution timescales between different metallicities.}
     \label{fig:histnovatime}
\end{figure*}

Figure~\ref{fig:histnovatime} shows the delay-time distributions of our \zml\ and \zmu\ metallicity cases, coloured by the accretor and donor stellar types. Delay time distributions such as those shown can be convolved with SFR histories to make predictions about nova populations in specific stellar environments\footnote{This process is discussed in more detail in K21.}. When considering the impact of varying metallicity in real stellar environments, these delay-time distributions are therefore of the highest importance. The results of convolving our data with age-metallicity-SFR relations are presented in Section~\ref{sec:results,galaxies}.

Reducing metallicity leads to reduced stellar life times. This is clearly seen in Figure~\ref{fig:histnovatime} when considering the stellar type of the donor star. The onset of significant contributions from FGB donor stars occurs by 3 Gyr at \zmu, compared to 1.5 Gyr at \zml. The earliest significant nova production -- primarily driven by TPAGB donors onto massive O/Ne WDs -- commences at around 100 Myr at \zml, compared to to around 50 Myr at \zmu. MS donors essentially disappear from the delay-time distribution at \zml\ by 5 Gyr, but remain minor sources of novae throughout the 15 Gyr simulation time at \zmu. Below \Z=\tento{-3} LMMS donor contributions to novae with delay-times $>$5 Gyr begin to drop noticeably and contributions from TPAGB donors increase.

In Section~\ref{sec:results,metalicity_sub:aggprod}, we discuss how lower metallicities lead to significantly higher nova production from $<1$ M\solar\ WDs at lower metallicites. This is reflected in the distribution of accretor compositions, particularly for the latest delay-times. At \zml, C/O contributions make up 40 per cent of all novae and over 80 per cent of novae beyond 10 Gyr. This is starkly different than the case at \zmu, where only 25 per cent of all novae are C/O WDs, and O/Ne WDs make up almost 60 per cent of all novae after $>$ 10 Gyr.

The shorter main-sequence lifetimes of low metallicity donor stars lead to a more prompt introduction of FGB donor contributions, producing delay-time distributions that have systematically higher contributions between 1-3 Gyr from low metallicity systems. This results in low-metallicity delay-time distributions tending to have a more gradual transition into the slowly decaying late-time tail of the distribution. Despite significant variation in the number of novae produced between low and high metallicities (see Figure \ref{fig:zsumtab}), this enhanced nova production is experienced over such a diverse set of channels that there is little change to the ratios of late versus early delay-time novae.

Surprisingly, the same appears to be true regarding relative contributions of different donor stellar types. Only minor changes are observed, with FGB, TPAGB, and HG donors showing enhanced contributions at the lowest metallicities, while LMMS and MS donors show reduced contributions. It should be emphasised that at all metallicities we find that FGB and TPAGB donors remain dominant, contributing to over half of all nova events.

This slight shifting towards giant donors at low metallicity reflects our previous commentary on the giant-donor-driven low-mass WD nova systems that produce a significant chunk of low metallicity novae. However, it is clear that the increase in nova productivity is not driven by this exclusively, supporting our previous assertion that the effect of metallicity on the final remnant mass is also of importance when considering the overall effect of metallicity on nova production.

\section{Predicted distributions and rates of novae in the Galaxy and M31}
\label{sec:results,galaxies}

In this section we present our predictions for the current nova rates and distributions of nova properties for the Milky Way Galaxy and M31. The process of forming these predictions is described in Section~\ref{sec:methods} and in K21.

Galactic nova populations are highly relevant to our understanding of the chemical evolution of our Galaxy, having bearing on problems such as the production of Galactic lithium \citep{izzo2015,tajitsu2015,starrfield2020} and chemical enrichment by type Ia supernovae. However, observationally estimating the Galactic nova rate is made difficult by complications such as extinction caused by interstellar dust, source crowding, systematic biases caused by our place in the Galaxy, and the selection functions of the surveys in which novae are detected \citep{shafter2001,darnley2006,shafter2017,dellavalle2020,de2021}. These issues are mitigated substantially by observing M31, where it is far easier to estimate completeness and account for a wide range of observational systematic uncertainties.

\subsection{The Milky Way}

\subsubsection{The Galactic nova rate}

We estimate a total Galactic nova rate of approximately 33 novae per year. There have traditionally been two observational approaches to estimating nova rates in the MW. One method uses small numbers of bright Galactic novae detected using optical instruments, combining this with completeness models that are highly sensitive to our understanding of Galactic interstellar dust. Estimates using this method typically arrive at numbers ranging from approximately 30-80 novae per year \citep{shafter1997,hatano1997,shafter2017,ozdonmez2018}. The second method relies on data from extra-galactic novae to extrapolate what the Galactic nova rate should be, based on our -- incomplete -- understanding of the differences between our own Galaxy's history and structure and those of the external galaxies \citep{dellavalle1994,shafter2000,darnley2006}. This method arrives at systematically lower nova rate estimates, ranging from approximately 20-40 novae per year \citep{ciardullo1990,
dellavalle1994,darnley2006}.

Recently, \cite{de2021} estimated a Galactic nova rate of $43.7^{+19.5}_{-8.7}$ novae per year based on 12 novae detected by Palomar Gattini-IR \citep[PGIR,][]{moore2019,de2020} in the near-infrared, where extinction caused by interstellar dust is far less prevalent. Figure 10 of \cite{de2021} presents a nice summary of the current state of the field regarding Galactic nova rates.

This work's estimate of 33 novae per year\footnote{Note that we do not attempt to account for any of the observational biases and selection criteria that plague observational Galactic rate estimates.} can be regarded as consistent with observational estimates deriving from extra-Galactic nova data. Relative to observational estimates deriving from Galactic optical nova data, as well as the \cite{de2021} estimate, our result can be considered an under-prediction, but not an excessive one. However, it should be noted that the nature of our methods means that our nova rates are inherently sensitive to modelling choices. This is both a strength and a weakness of this methodology. It is likely that our predicted Galactic nova rate could vary by as much as 30 novae per year under different combinations of physical assumptions.

\subsubsection{Predicted properties of Galactic novae}

In our adopted age-metallicity-SFR relation, shown in Figures~\ref{fig:sfrMW} and~\ref{fig:zvtimeMW}, the current SFR is assumed to be zero in the thick disk and the halo, with most new stars born in the thin disk and a small fraction (<10 per cent) born in the bulge. These components also differ significantly in their evolution histories. The halo represents the oldest, most metal-poor component, followed by the thick disk. The metallicity in the bulge rises very rapidly at early times, forming few metal-poor stars, while the metallicity in the thin disk rises more gradually. We find that 70 per cent of the current novae in the MW occur in the thin disk, 24 per cent in the bulge, 5 per cent in the thick disk, and 1 per cent in the halo. 

\begin{figure*}
\centering
\includegraphics[width=0.8\textwidth]{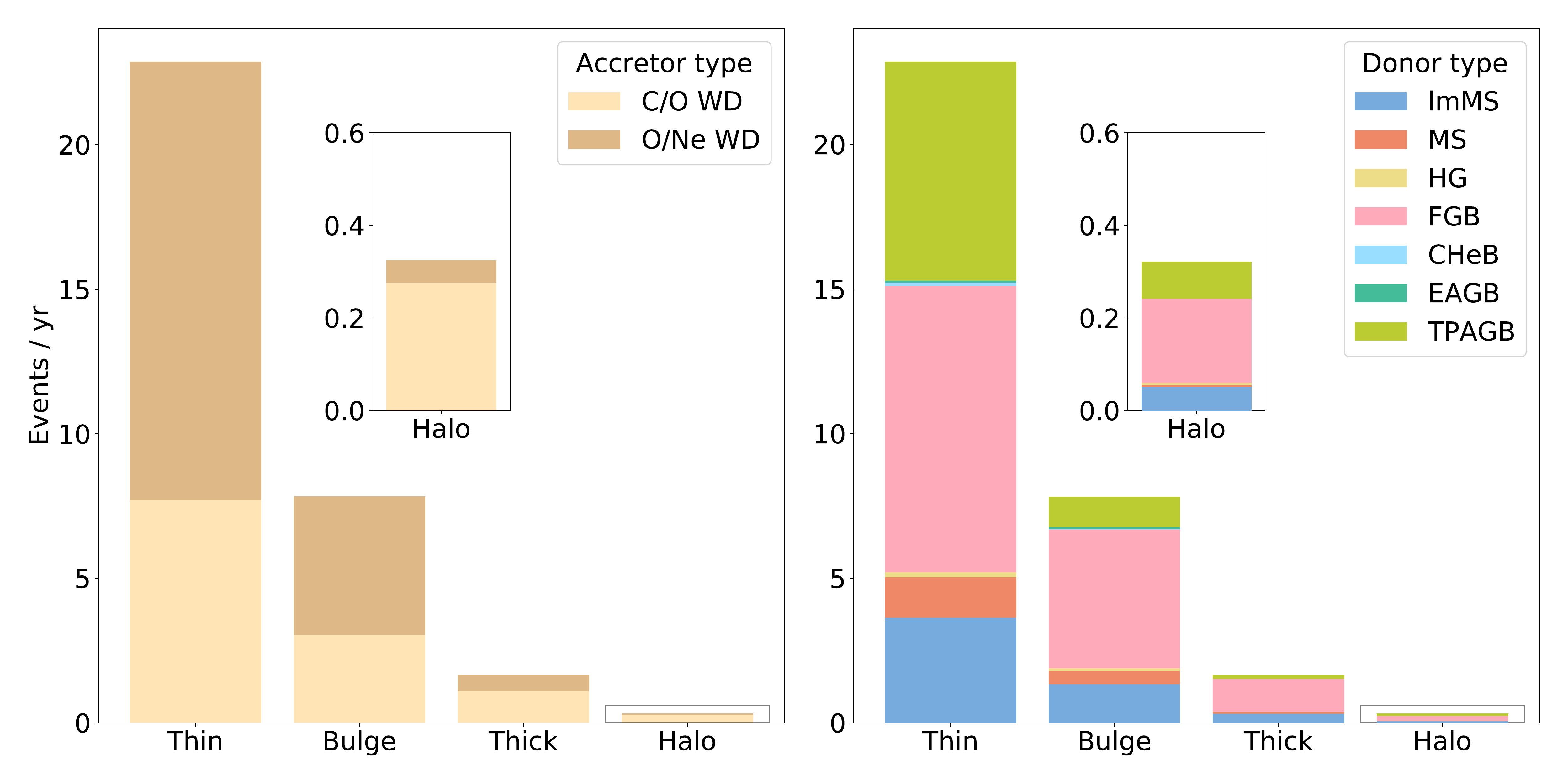}
\caption{Nova contributions to the predicted current Galactic nova rate for each Galactic component. Colouration breaks down each component further into the accretor (left) and donor (right) stellar types (see Table~\ref{tab:evotags} for stellar type glossary). The total predicted Galactic nova rate is approximately 60 novae per year.}
\label{fig:stbarMW}
\end{figure*}

Figure~\ref{fig:stbarMW} shows the break-down of accretor and donor stellar types of novae at the current galaxy age (13 Gyr) for each Galactic component. There are clear systematic differences in the relative importance of the donor and accretor stellar types. FGB donor systems are found to be important contributors of novae in all four Galactic components, reflecting the prevalence of FGB donors in novae beyond 3 Gyr delay-times at all metallicities (see Section~\ref{sec:results,metalicity_sub:dtd}). TPAGB donor systems present an interesting case, as so much of their production occurs within the first Gyr, but at low metallicities a non-negligible number of late-time TPAGB-driven novae appear, typically making up around 20 per cent of the nova production. Thus the relative importance of TPAGB donor systems is sensitive both to the metallicity history and the age of a stellar population. The significant TPAGB contributions to thin disk and bulge novae (approximately 35 and 15 per cent, respectively) are to be expected, as both the thin disk and bulge are currently forming stars and so early-delay-time novae are still occurring here. However, we also find that TPAGB-donor driven novae account for up approximately 25 per cent of halo novae. In fact, the \textit{relative} importance of TPAGB donors is higher in the halo than in the bulge. This feature is entirely driven by the exclusively metal-poor nature of the halo. The thick disk, which has more metal-rich systems than the halo, has a far lower relative contribution from TPAGB donor systems.

Regardless of the Galactic component under consideration, novae with giant (FGB and TPAGB) donor stars are found to contribute to more than half of the currently observable nova population. The importance of giant donor stars to novae has previously been discussed in the context of recurrent and symbiotic novae \citep[e.g.,][]{mikolajewska2010,darnley2012,darnley2019}. Known examples of these novae make up less than 20 of the hundreds of detected Galactic novae, and there exists to date no robust observational estimate of the relative importance of giant donor stars in Galactic nova systems. Figure 10 of K21 indicates that roughly half of all nova eruptions involving giant donors have recurrence times greater than 100 years. In the absence of a comparable observational estimate for the fraction of giant- versus main sequence-donor novae in the Galaxy, we can only state that our results indicate that nova systems with giant donors produce significantly more novae than current Galactic observations suggest.

The distribution of WD compositions (C/O vs O/Ne) is found to be dominated by the age of the stellar population. The actively star-forming thin disk and bulge exhibit significantly higher nova contributions from O/Ne WDs (66 and 61 per cent respectively), while O/Ne WDs make up a much smaller fraction of the novae in the thick disk and the halo (32 and 16 per cent respectively).

\begin{figure*}
\begin{subfigure}{0.9\columnwidth}
    \centering
    \includegraphics[width=\textwidth=1]{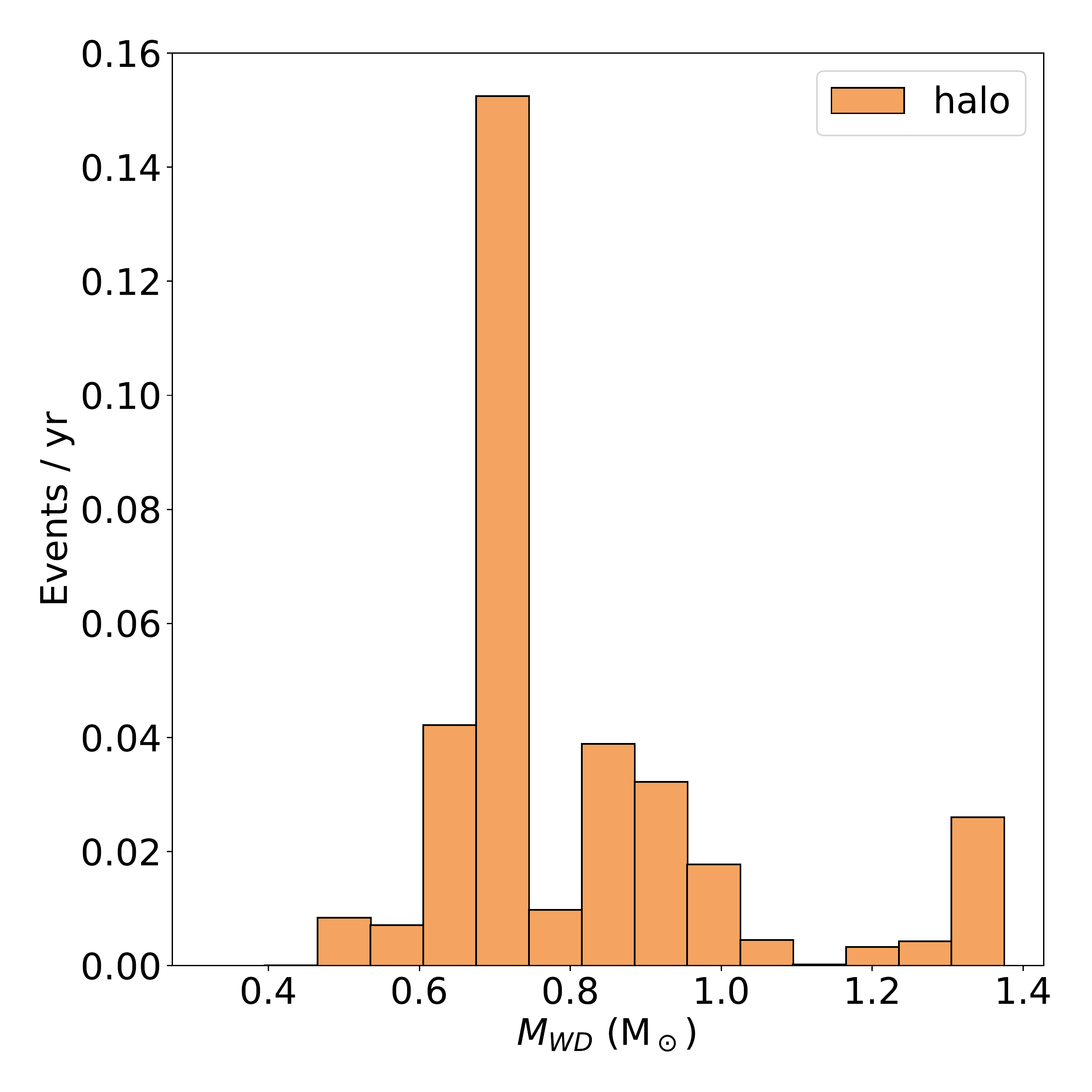}
    \caption{Halo}
\end{subfigure}%
\begin{subfigure}{0.9\columnwidth}
    \centering
    \includegraphics[width=\textwidth=1]{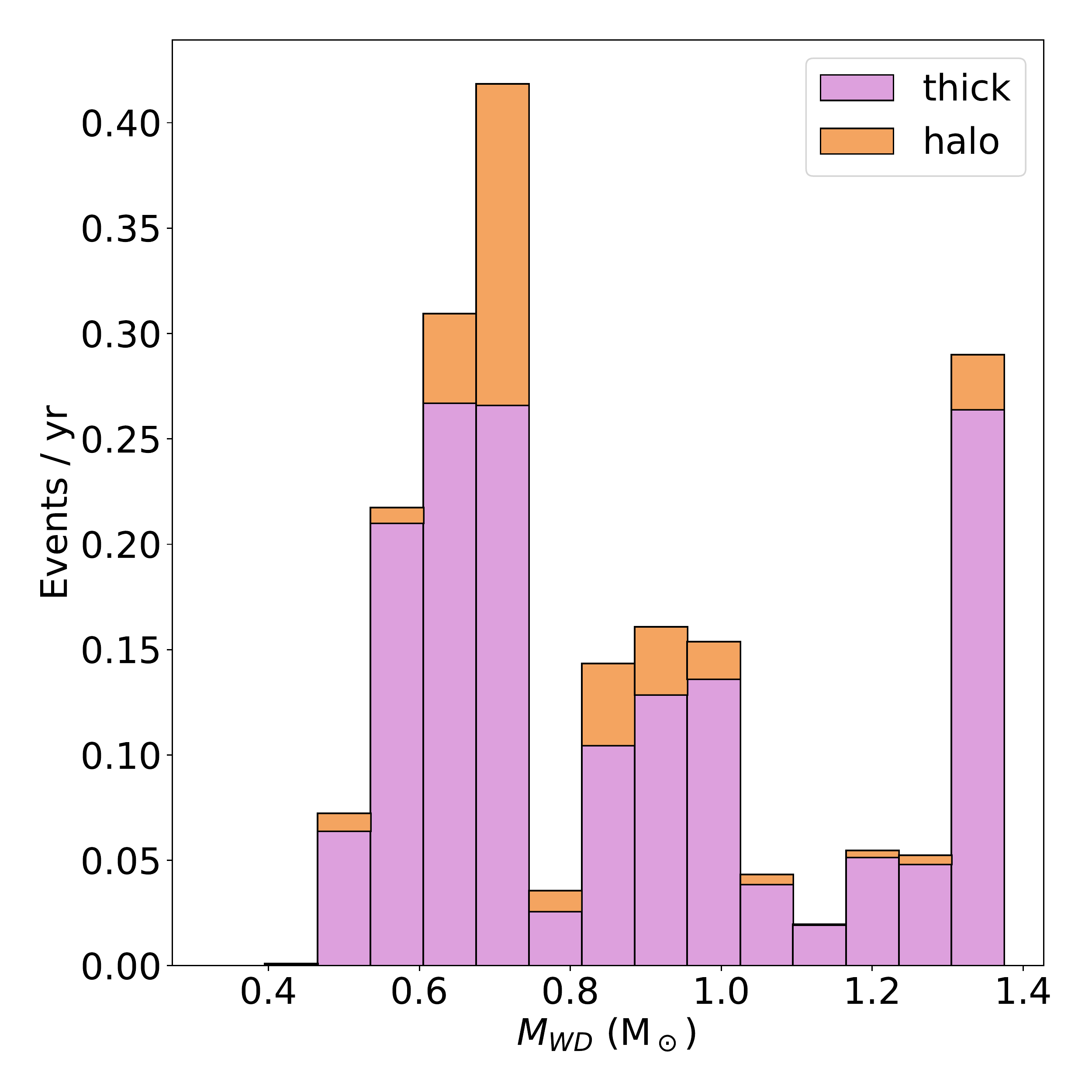}
    \caption{Thick disk}
\end{subfigure}

\begin{subfigure}{0.9\columnwidth}
    \centering
    \includegraphics[width=\textwidth=1]{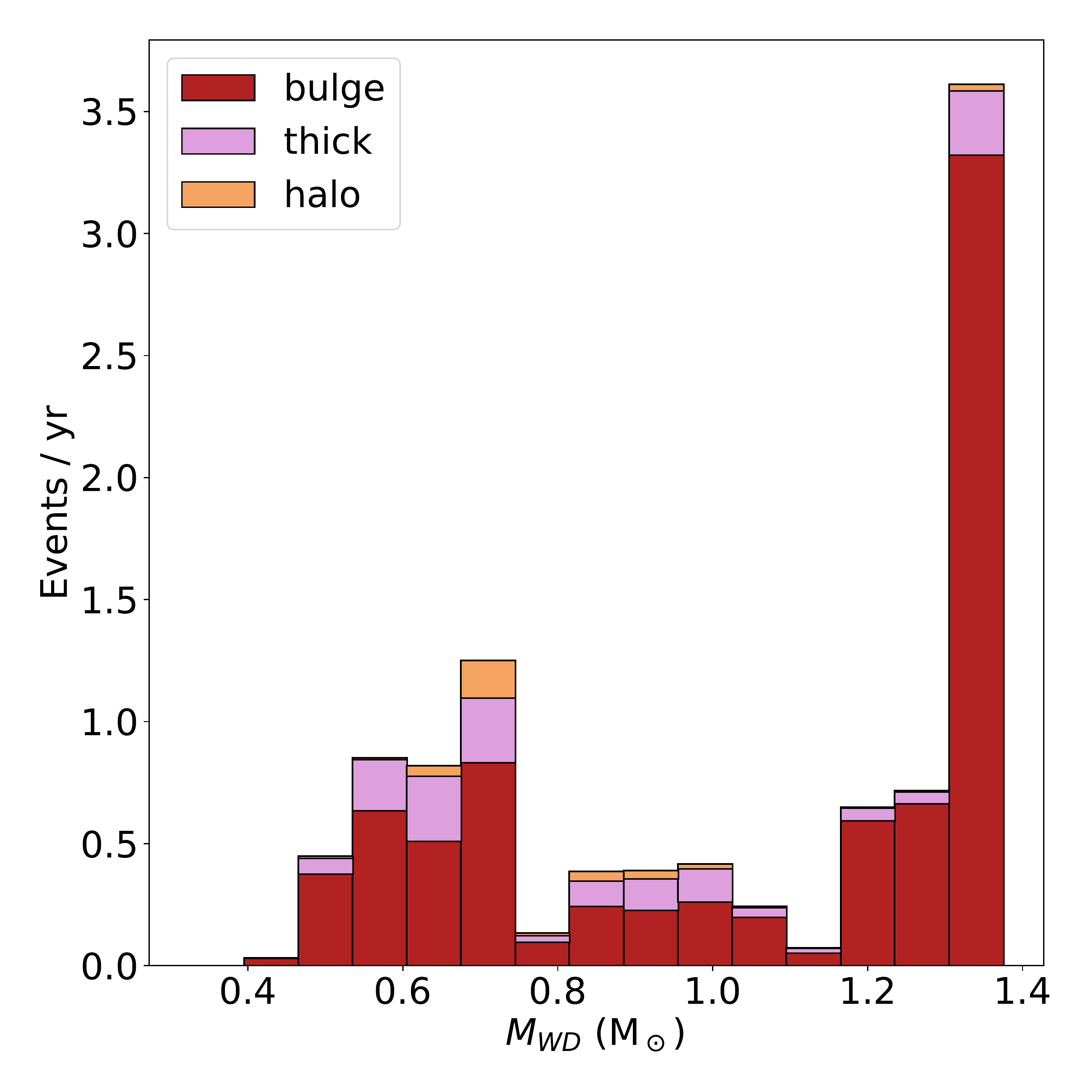}
    \caption{Bulge}
    
    \label{fig:mwdsliceMWbulge}
\end{subfigure}%
\begin{subfigure}{0.9\columnwidth}
    \centering
    \includegraphics[width=\textwidth=1]{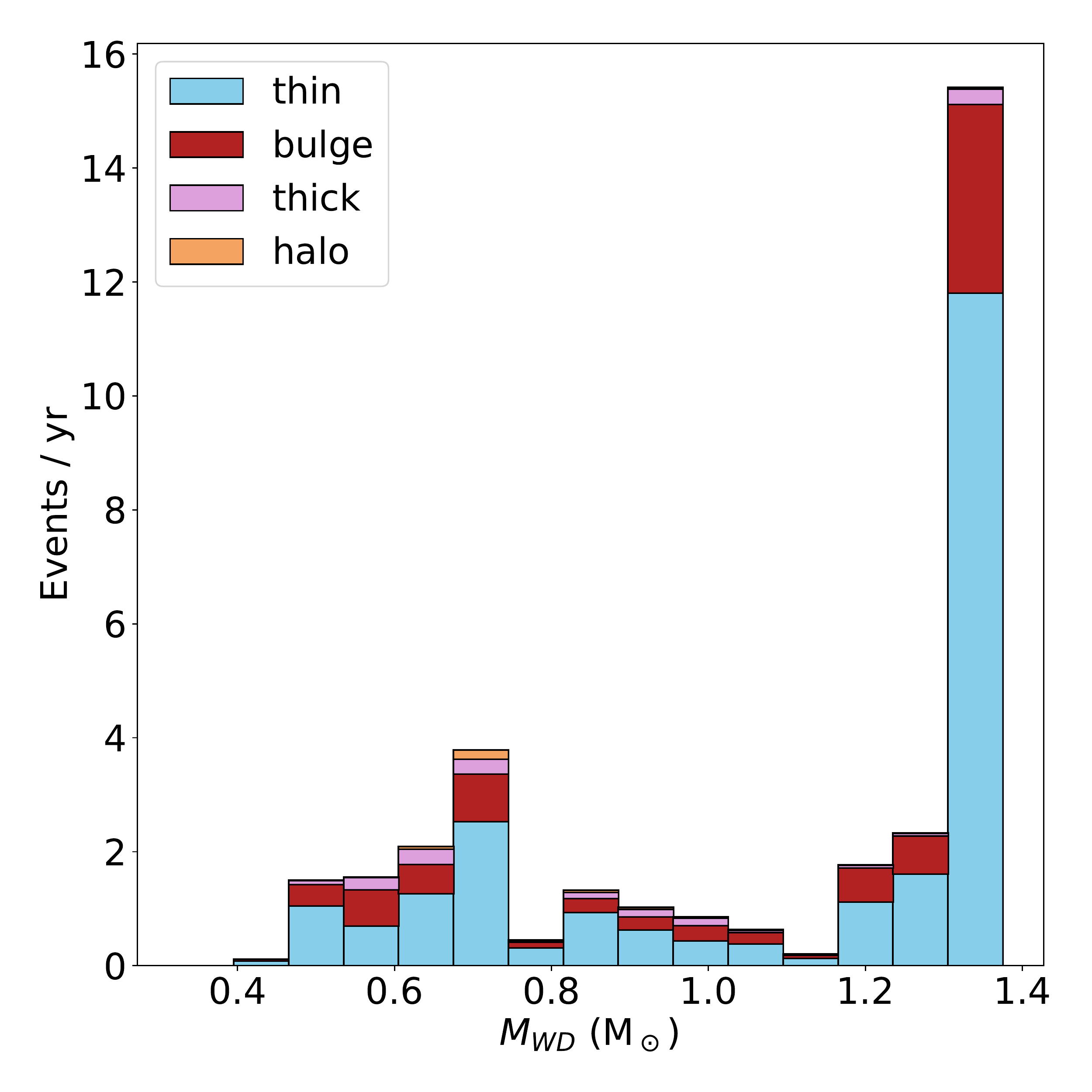}
    \caption{Thin disk}
    \label{fig:mwdsliceMWthin}
\end{subfigure}
\caption{Nova WD mass distribution predictions for each Galactic component. The predicted full Galactic distribution can be observed in the bottom right panel as the sum of the contributions of each component. The similarity between the thin disk and bulge (both of which are currently undergoing star formation in our model) distributions highlights the dramatic effect that the presence of young nova systems can have on nova populations.}
\label{fig:mwdsliceMW}
\end{figure*}

Figure~\ref{fig:mwdsliceMW} shows the predicted distributions of WD masses in each Galactic component, with the full Galactic distribution presented in Figure~\ref{fig:mwdsliceMWthin}. This distribution reveals a nova population dominated by high mass WDs occurring in the thin disk and bulge. Comparison between the Galactic components reveals this high-mass peak to be strongly associated with young stellar systems. The stellar populations where star formation has ceased lack this feature, with the halo predicted to contain very few novae with high mass WDs, instead peaking strongly around 0.7 M\solar, and the thick disk displaying comparable nova contributions in peaks around \Mwd\ = 0.7, 0.9, and 1.3 M\solar. The bulge and the thin disk, on the other hand, have very similar distributions of \Mwd\, characterised by the large peak around the most massive WDs.

The results of these predicted nova distributions are in principle comparable with Galactic observational estimates of nova distributions. Figure 2 of \cite{shara2018} presents an observationally derived WD mass distribution using Galactic nova light curves which peaks around \Mwd=1.15 M\solar. Put bluntly, there is no resemblance between the predicted WD mass distribution shown in Figure~\ref{fig:mwdsliceMWthin} and this distribution. While it should be acknowledged that observational selection biases -- as well as the assumptions necessarily employed in \cite{shara2018}, which relies on its own set of physical models -- may play a role here, given the magnitude of the discrepancies it is far more likely that the adopted combination of physical choices simply do a poor choice at reproducing the WD distribution. This is not overly concerning. Given the large parameter space -- and associated uncertainties -- in binary stellar evolution, it would have been surprising if our initial `best-guess' of binary physics were able to reproduce observations. The important result here is that predictions of the WD distribution are likely to be very sensitive to physical assumptions, and therefore have the potential to be an important tool in identifying good combinations of binary physics.

\begin{figure}
\centering
\includegraphics[width=0.8\columnwidth]{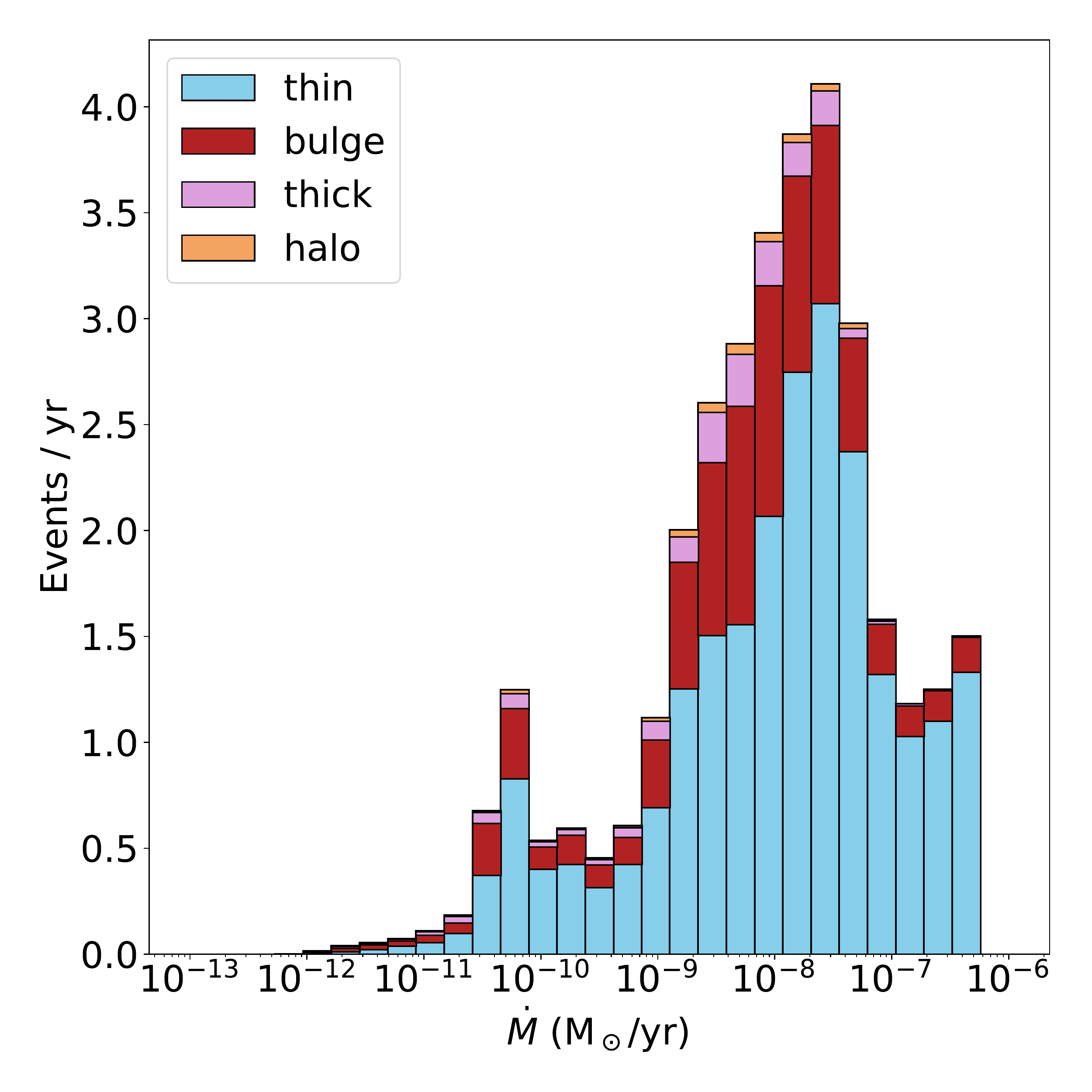}
\caption{Predicted current Galactic nova accretion rate distribution. The distributions of each Galactic component show only minor variations in morphology, see main text for details.}
\label{fig:mdotsliceMW}
\end{figure}

Figure~\ref{fig:mdotsliceMW} presents the predicted distribution of WD accretion rates. There is relatively little variation between the different Galactic components, all of which show the same basic profile. The highest accretion rates ($>$ \tento{-7} M\solarperyr) are almost absent in the halo and most heavily emphasised in the thin disk, but the overall structure remains, with a minor peak around \tento{-10} M\solarperyr\ and the main peak around \tento{-8} M\solarperyr. The lack of significant variation can be at least partially attributed to the prevalence of FGB donors across all Galactic components (see Figure~\ref{fig:stbarMW}).

Figure 3 of \cite{shara2018} presents an observationally-derived distribution of nova accretion rates for Galactic novae. There are, once again, significant differences between this distribution and our predicted distribution shown in Figure~\ref{fig:mdotsliceMW}. Figure 3 of \cite{shara2018} shows a distribution peaking around \tento{-9} M\solarperyr, while our prediction peaks a full order of magnitude higher at \tento{-8} M\solarperyr. The observed distribution shows no novae beyond \tento{-7} M\solarperyr, while our prediction shows significant contributions up to around \timestento{5}{-7} M\solarperyr. These discrepancies are significant, although they are minor compared to those seen in the \Mwd\ distributions. Similarly to the \Mwd\ distribution, we conclude that the mass accretion rate distribution also shows promise as a statistic to compare different physical assumptions to observations. However, it should be noted that this practical goal is complicated by the reliance of studies that derive these distributions from nova light curves on detailed theoretical nova models. An investigation dedicated to the degree of uncertainty in the observed distributions of nova WD masses and accretion rates resulting from different physical assumptions and model sets would be extremely useful.

\begin{figure*}
\begin{subfigure}{0.9\columnwidth}
    \centering
    \includegraphics[width=\textwidth=1]{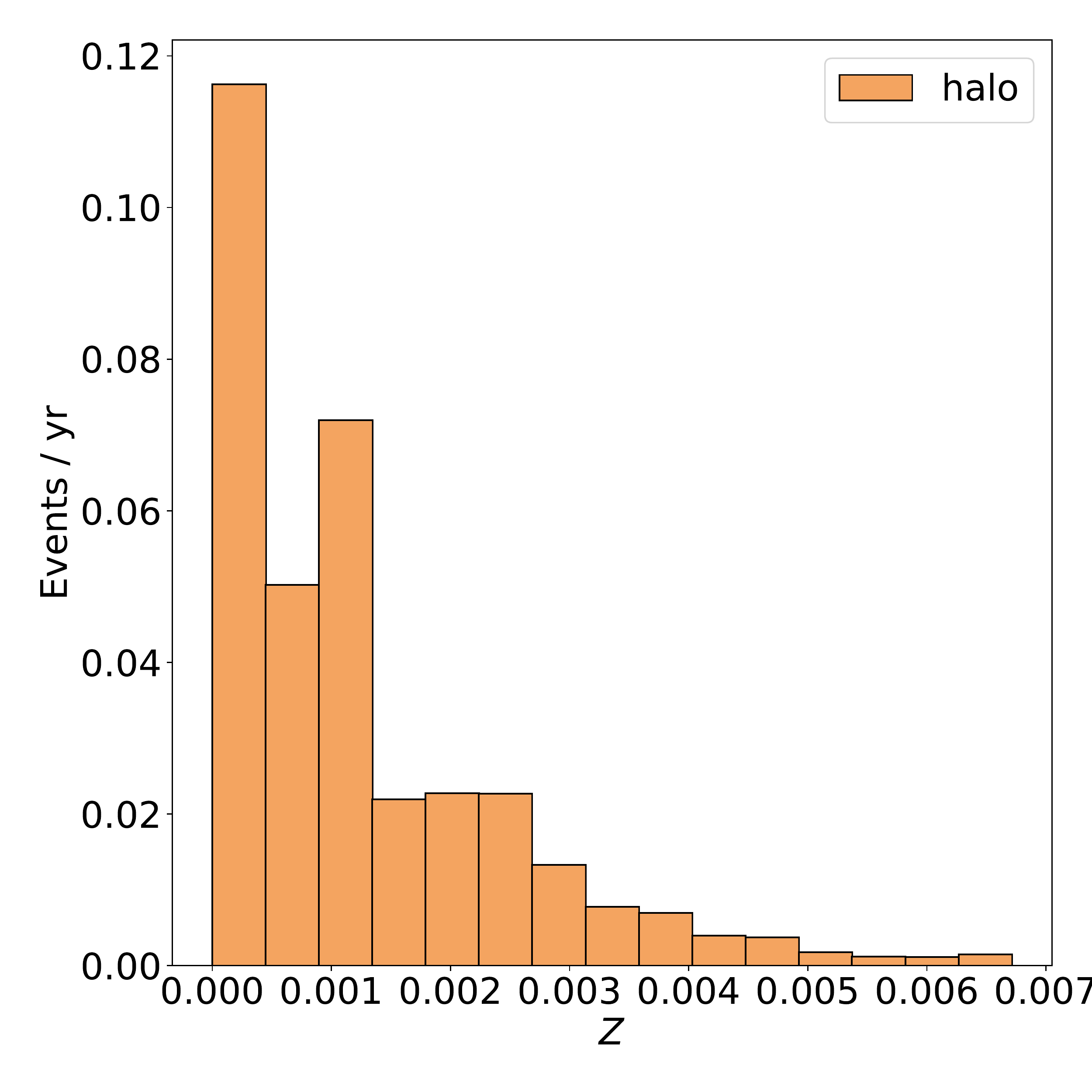}
    \caption{Halo}
\end{subfigure}%
\begin{subfigure}{0.9\columnwidth}
    \centering
    \includegraphics[width=\textwidth=1]{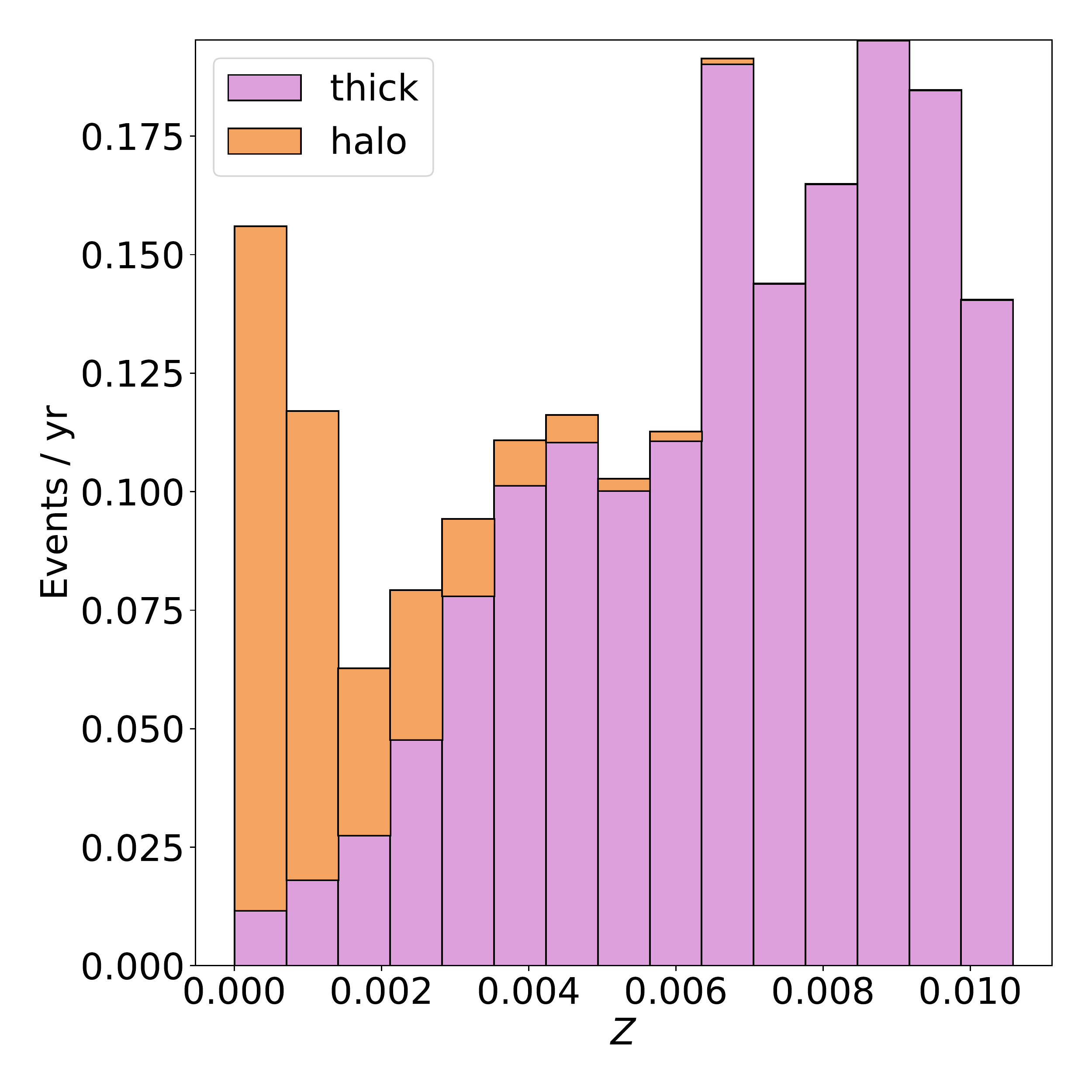}
    \caption{Thick disk}
\end{subfigure}

\begin{subfigure}{0.9\columnwidth}
    \centering
    \includegraphics[width=\textwidth=1]{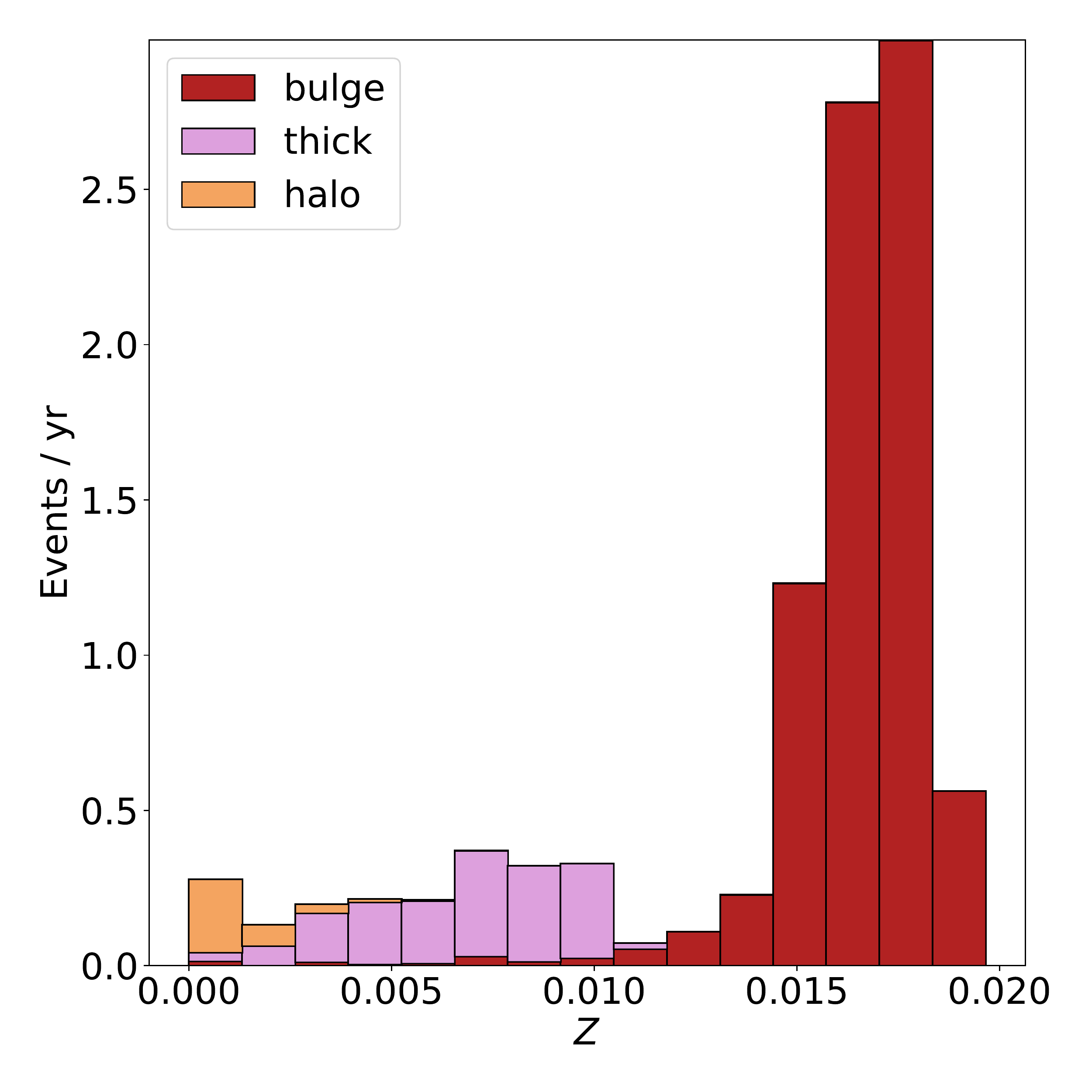}
    \caption{Bulge}
\end{subfigure}%
\begin{subfigure}{0.9\columnwidth}
    \centering
    \includegraphics[width=\textwidth=1]{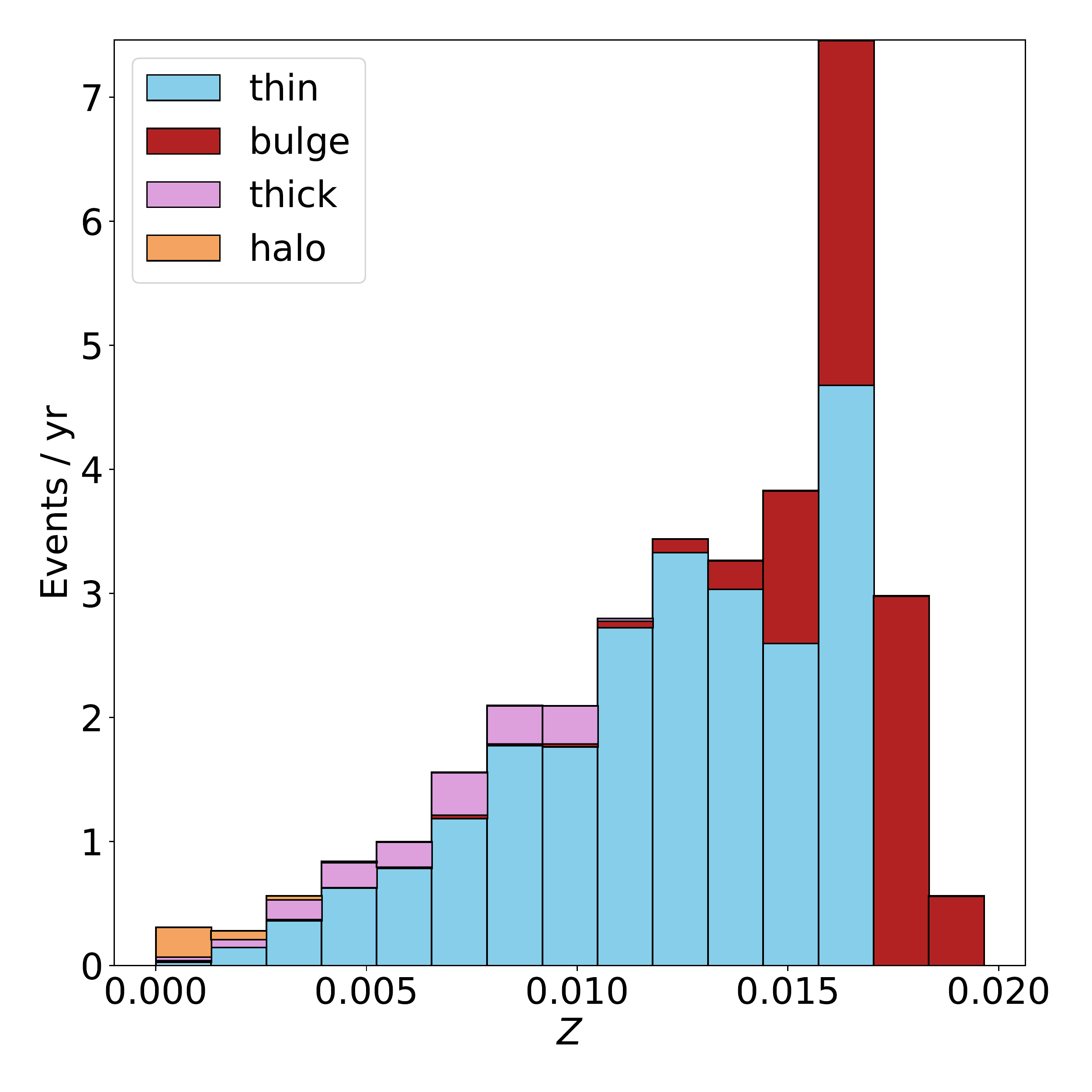}
    \caption{Thin disk}
\end{subfigure}
\caption{Current nova metallicity distribution predictions of each Galactic component. The full Galactic distribution can be observed in the bottom right panel as the sum of the contributions of each component. Significant variation is present between these different stellar environments, with the exception of the thin disk and the bulge.}
\label{fig:ZsliceMW}
\end{figure*}

Finally, the predicted metallicity distribution of novae for each Galactic component is shown in Figure~\ref{fig:ZsliceMW}. The degree of variation within the different panels of this figure is remarkable. The halo distribution decreases rapidly towards higher metallicities, with most halo novae characterised by metallicities less than \timestento{2}{-3}. In fact, despite the extremely low number of halo novae ($\approx$ 0.3 novae per year), the halo actually dominates the population of novae at the lowest metallicities. The thick disk has a broad, increasing metallicity profile that cuts off sharply at $Z\approx0.01$, corresponding to the metallicity at the time of the sudden cut-off in star formation at 3 Gyr shown in Figure~\ref{fig:sfrMW}. The bulge exhibits a narrowly peaked distribution, with almost all novae occurring in systems with metallicities between 0.015 and 0.02. The thin disk profile is qualitatively similar to the thick disk, having a broad distribution that rises steadily with increasing metallicity, peaking just beyond \Z=0.015.

The total Galactic distribution peaks strongly where the end of the thin disk distribution partially aligns with the peak of the bulge. Its overall shape closely follows that of the thin disk. It is clear that the extremes of the metallicity distribution are expected to be dominated by bulge and halo novae, however there is an important caveat to keep in mind when assessing Figure~\ref{fig:ZsliceMW}. In nature, there are composition variations within the different Galactic components, with star formation occurring at a range of metallicities simultaneously. However, our age-metallicity-SFR model does not not account for any scatter in \Z, instead simply taking the value of \Z\ that represents the average metallicity for that component.

\subsection{M31}

\subsubsection{M31 nova rate}

\begin{figure}
\centering
\includegraphics[width=1\columnwidth]{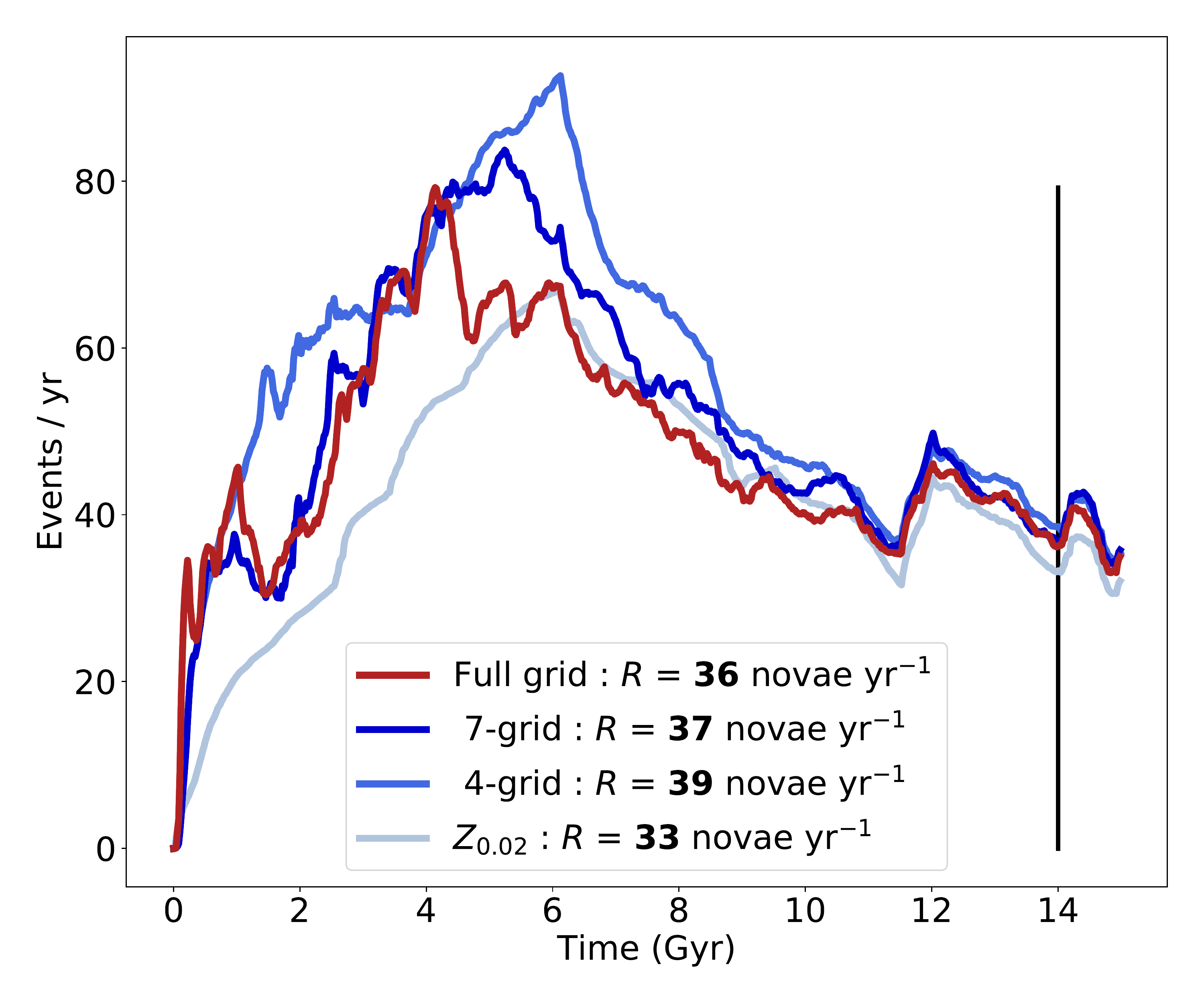}
\caption{Nova rate history in M31, computed using four different metallicity grids. The `full grid' uses data from every grid computed (Figure \protect~\ref{fig:zsumtab}). `7-grid' uses only the \zml, \tento{-3}, \timestento{2.5}{-3}, \timestento{5}{-3}, $0.01$, $0.015$, and $0.02$ grids. `4-grid' uses only the \zml, \tento{-3}, $0.01$, and $0.02$ grids, and `\Z$_{0.02}$' uses only the \Z=0.02 grid. The estimates of the current (t=14 Gyr, vertical black line) event rate $R$ are given in the legend.}
\label{fig:eventratehistorym31}
\end{figure}

M31's close proximity, favourable inclination angle, and relatively well understood observational systematics make it the best available laboratory for studying nova rates. \cite{darnley2006} present the most recent observational rate estimate of $65^{+15}_{-16}$ novae per year, with older works estimating nova rates between 20 and 50 novae per year \citep{hubble1929,arp1956,capaccioli1989,shafter2001}. Convolving our full, 16-point metallicity grid with the M31 age-metallicity-SFR relation described in Section~\ref{sec:methods}, we estimate a current nova rate of approximately 36 novae per year, very close to the \cite{darnley2006} estimate.

Figure~\ref{fig:eventratehistorym31} presents four event rate histories, each computed using the same age-metallicity-SFR relation but with different metallicity grids. The `full grid' history uses all 16 metallicities (see Table~\ref{tab:zsumtab} and Figure~\ref{fig:zsumtab}), the `7-grid' history uses only the \zml, \tento{-3}, \timestento{2.5}{-3}, \timestento{5}{-3}, $0.01$, $0.015$, and $0.02$ grids, the `4-grid' history uses only the \zml, \tento{-3}, $0.01$, and $0.02$ grids, and the `\Z$_{0.02}$' history uses only the \Z=0.02 grid. The \Z$_{0.02}$ history is representative of the simplified case computed in K21\footnote{Note that K21 estimated an event rate of $41 \pm 4$ novae per year. This differs from the $Z_{0.02}$ rate of 33 novae per year presented in Figure \ref{fig:eventratehistorym31} due to the combined effect of a technical error present in K21, which caused the assumed 50 per cent binary fraction to be improperly implemented, and discrepancies between the assumed age of M31. See Section \ref{sec:methods} for details.}, which ignored metallicity effects.

The predicted \textit{current} nova rates for M31 are very similar between each of the 4 grids. The full grid predicts 36 novae per year, the 7-grid predicts 37, the 4-grid predicts 39, and the \Z$_{0.02}$ grid predicts 33. However, the event rate histories themselves show significant variation, particularly in the first 6 Gyr of M31's evolution. 

It might be expected that the \Z$_{0.02}$ grid would perform the most poorly when compared to the full metallicity grid, with no low-metallicity data to work with. However, this expectation is only partly borne out. Relative to the full grid event rate history, event rates are under-predicted for the first 6 Gyrs of M31's life, but beyond this point the \Z$_{0.02}$ grid is able to closely match the `full grid' prediction. This result reflects how important young nova systems are. Despite incorrectly predicting the production of ancient nova-systems, the \Z$_{0.02}$ history actually turns out to be a fairly reliable predictor of the event rates in M31 for the last 8 Gyrs.

However, Figure~\ref{fig:eventratehistorym31} also carries a cautionary message. The 4-grid metallicity systematically over-predicts the event rate for almost the entire evolution of M31, performing even worse than the \Z$_{0.02}$ grid. The 4-grid metallicity, with its poor sampling of the \Z<0.01 part of the parameter space, results in a large spike in nova production at around 6 Gyr, in addition to other deviations around 2 Gyr. It is primarily the late-time tail of the 6 Gyr feature that drives the lower level over-prediction for the remainder of M31's life.

Contrasting with the poor approximation of the 4-grid, the 7-grid set (which additionally includes results for \Z=\timestento{2.5}{-3}, \timestento{5}{-3}, and 0.015) does a far better job at approximating the full grid history for the entire lifetime of M31. The deviation around 5 Gyr is far smaller than that of the 4-grid, and elsewhere this set approximates the event history of M31 very closely. This is unsurprising in some ways, as this metallicity-set was selected by identifying the short-comings of the 4-grid set and then including additional metallicity grids where necessary.

It is clear that the results of convolutions between age-metallicity-SFR relations and grids of delay-time distributions that poorly resolve metallicity -- particularly in the \Z<0.01 metallicity regime -- should be treated with caution, as under-resolving can potentially result in serious systematic errors. However, the success of the 7-grid solution in approximating the full grid event history demonstrates the power of a carefully designed metallicity grid that is tailored to a specific age-metallicity-SFR relation.

\subsubsection{Predicted properties of novae in M31}

\begin{figure}
\centering
\includegraphics[width=0.8\columnwidth]{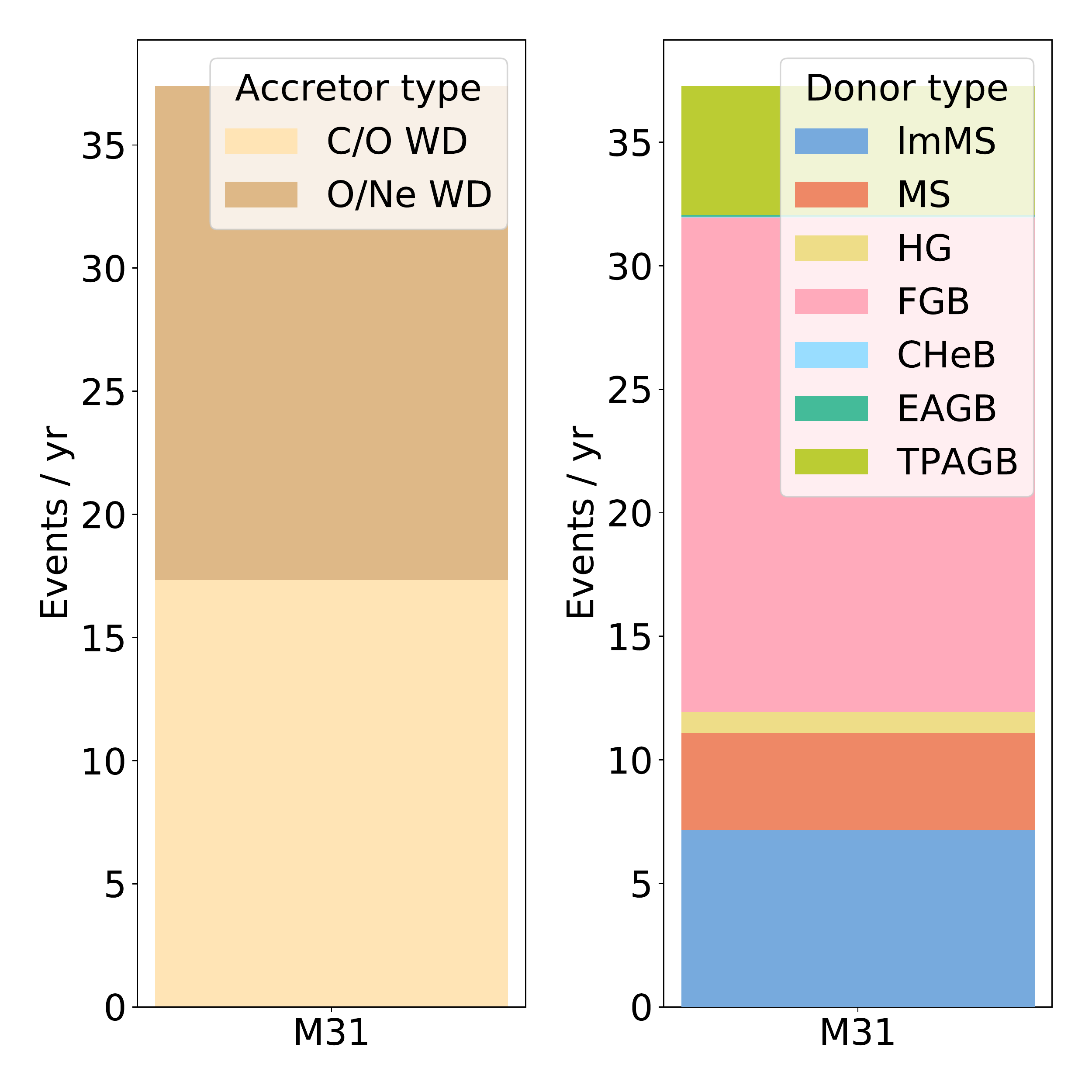}
\caption{Break-down of the accretor and donor stellar types (see Table~\ref{tab:evotags} for stellar type glossary) for M31 novae as they would appear today, analogous to Figure~\ref{fig:stbarMW}. Of all the different Galactic components, the M31 distribution most closely resembles the Milky Way's bulge.}
\label{fig:stbarM31}
\end{figure}

Figure~\ref{fig:stbarM31} presents the relative contributions of the different stellar types of M31 novae, and is analogous to Figure~\ref{fig:stbarMW}. The M31 distributions most closely resemble the bulge component in the MW. The most likely reason for this is that the age-metallicity-SFR model of both M31 and the Galactic bulge are characterised by an early peak in star formation. This leads to a larger proportion of old nova systems, which tend to more heavily favour C/O WD and FGB donor contributions over O/Ne WD and TPAGB contributions.

It is of interest to note that Figure \ref{fig:stbarM31} indicates that roughly two thirds of all currently observable novae in M31 are predicted to have giant (FGB or TPAGB) donors. This is significantly higher than the $30^{+13}_{-10}$ per cent fraction derived by \cite{williams2014,williams2016} through statistical analysis of M31 novae. The degree to which this discrepancy is due to possible biases against the identification of giant-donor novae \citep{williams2016} versus systematic biases caused by our physical assumptions is unclear.

\begin{figure}
\centering
\includegraphics[width=1\columnwidth]{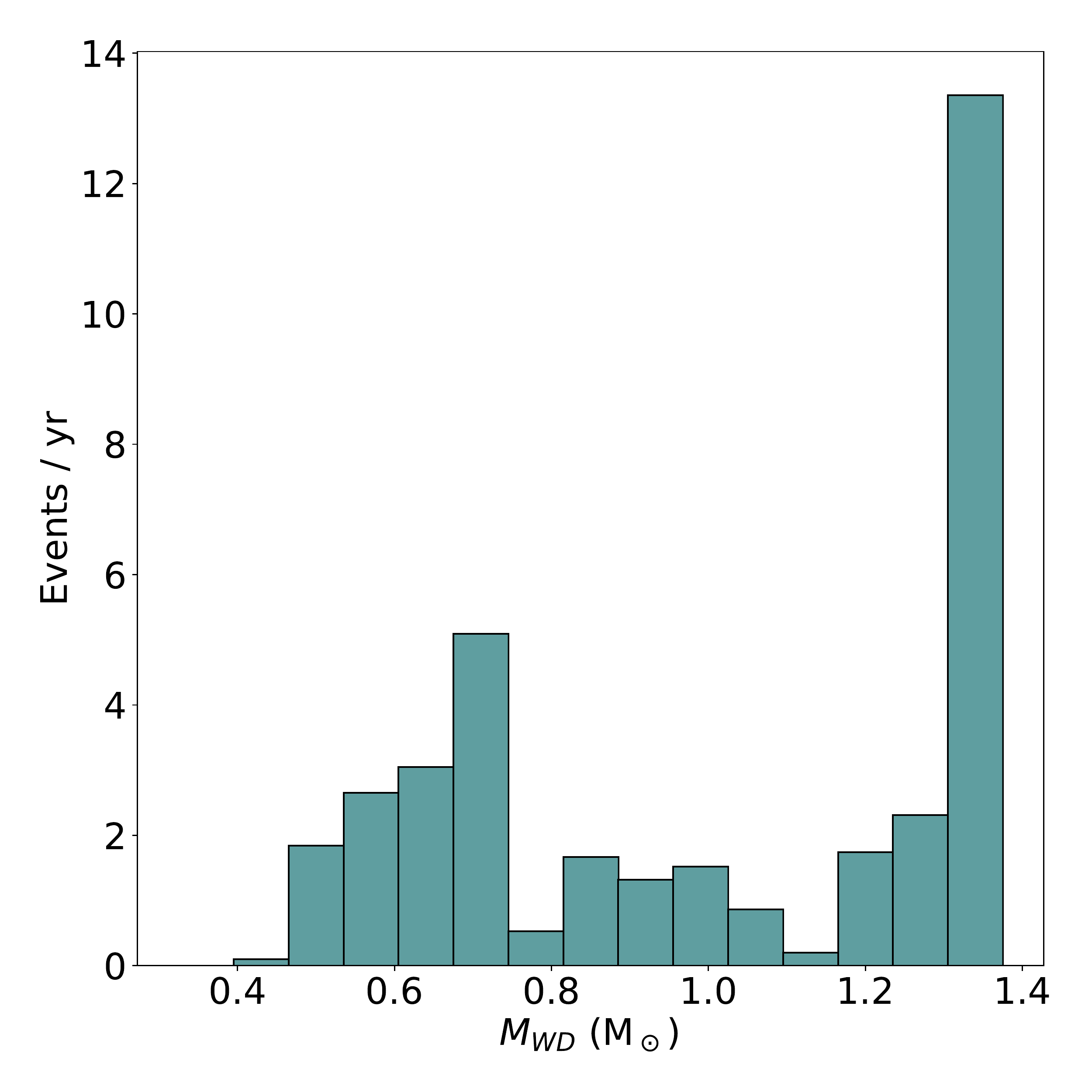}
\caption{Predicted current distribution of nova white dwarf masses in M31. This distribution is similar to that of the Milky Way (Figure~\ref{fig:mwdsliceMWthin}).}
\label{fig:mwdsliceM31}
\end{figure}

This similarity between the Galactic bulge and M31 can also be seen in the predicted distribution of WD masses shown in Figures~\ref{fig:mwdsliceMW} and~\ref{fig:mwdsliceM31} for the MW and M31, respectively. Note that in this case, the distribution for the Galactic thin disk also very closely resembles M31, as does the Galactic distribution as a whole. Given how different the SFR histories for M31 and the MW are, this similarity strongly reinforces our previous assertion that whether or not a stellar environment is undergoing (or has recently undergone) star formation is extremely important when considering the WD mass distribution of novae. Neglecting the metallicity evolution of M31 entirely and using only \Z=0.02 data, analogous to the method used to approximate M31 in K21, results in minor changes to the distribution, increasing the importance of the most massive (>1.25 M\solar) WDs and reducing the importance of events at lower (0.6-1 M\solar) white dwarf masses.

Just as we find minimal discrepancies between the shape of the accretion rate distributions in different Galactic components, we also find no significant variation between the accretion rate of the MW and that of M31 (not shown). It appears that the prevalence of FGB donors in late delay-time novae make distributions of nova accretion rates extremely poor tools for discriminating between different stellar environments.

\begin{figure}
\centering
\includegraphics[width=1\columnwidth]{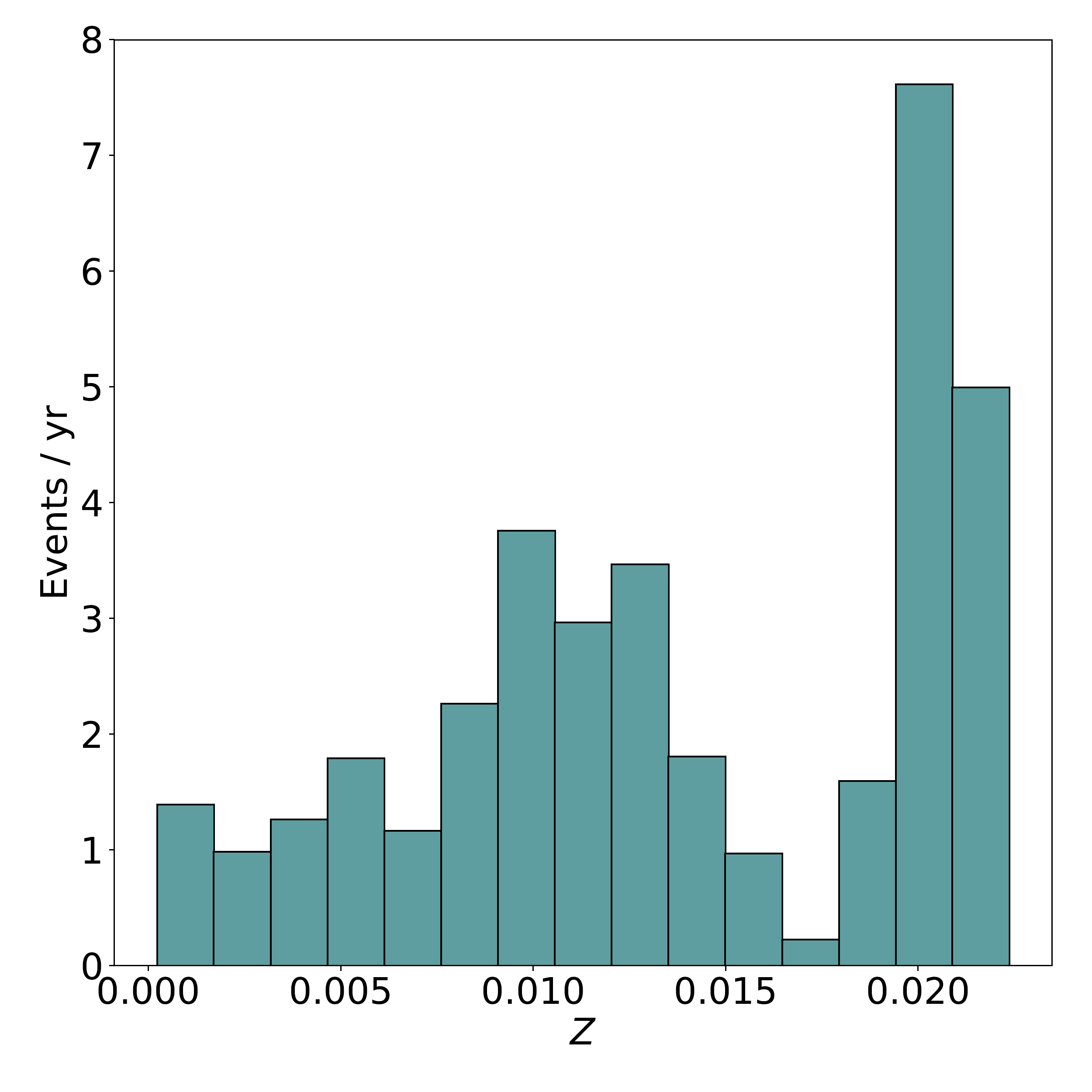}
\caption{Predicted current distribution of nova metallicities for M31. The trough around \Z=0.017 is caused by the lapse in star formation in our M31 age-star formation rate model between 7.5 and 9 Gyr (see Figure~\ref{fig:SFRvtimeM31}).}
\label{fig:ZsliceM31}
\end{figure}

The metallicity distribution of the novae, however, may prove to be an excellent tool for such a task. Figure~\ref{fig:ZsliceM31} shows the predicted metallicity distribution of novae in M31, and it is dramatically different from any of the Galactic components. 

In M31 we see vaguely bimodal metallicity distribution with a broad peak from \Z=0.1-0.13, a trough at \Z=0.017, and a second, sharper peak at \Z=0.02. Figure~\ref{fig:SFRvtimeM31} shows that between 7.5 and 9 Gyr, M31 -- at least in the \cite{williams2017} model -- experienced a significant lapse in star formation. This period of time matches well with the \Z=0.017 location of the trough in the metallicity distribution seen in Figure~\ref{fig:ZsliceM31}. Figure~\ref{fig:ZsliceM31} suggests a very broad distribution of nova metallicities in M31, resulting from the combination of significant ancient star formation and a lower level of more recent star formation.

In some ways, the details of Figure~\ref{fig:ZsliceM31} are unimportant. The age-metallicity component of our age-metallicity-SFR model for M31 is simple to the point of being arbitrary, and so it is questionable whether Figure~\ref{fig:ZsliceM31} represents an accurate picture of the true metallicity distribution of novae in M31. What \textit{is} important is that the metallicity distributions of novae change significantly between different star forming environments. This gives real hope to the idea that distributions of nova metallicities may be used as a probe of the evolution of distant stellar environments.

\section{Discussion}
\label{sec:discussion}

\subsection{Metallicity dependence of detailed nova properties}

\label{sec:chen2019disc}
In this work, we study the effect that metallicity-dependent stellar evolution has on nova populations, and neglect effects due to the metallicity-dependence of detailed nova properties such as \Mig, $\eta$, or the accretion limits. Here, we assess the probable effect that accounting for these additional dependencies would have on our predictions for Galactic novae.

\cite{chen2019} present detailed nova calculations for the \zml\ and $Z=0.02$ cases over a range of WD masses, in addition to varying mixing parameters. When comparing these two extremes of the metallicity distribution, the critical ignition masses, which directly effect nova rates, increase slightly when metallicity is reduced, with the \zml\ case typically being around 0.2 dex higher than the \Z=0.02 case. However, accretion efficiencies increase dramatically as metallicity is decreased \citep[see Figures 9-10,][]{chen2019}, and the lower accretion limit for steady burning (which provides a ceiling for the nova regime\footnote{The nova regime is the region in the \Mwd-\Mdot\ parameter space where novae may occur.}) reduces by 0.3-0.5 dex (Figure 1, \cite{chen2019}). It is noteworthy that varying the mixing parameter from 0 to 0.25 produces greater variation in the critical ignition masses and accretion efficiencies than varying the metallicity, and that the WD mass and accretion rate remain the most important considerations when calculating these key properties.

Our current work relies on solar-metallicity input nova models, using critical ignition masses, accretion efficiencies, and accretion regimes drawn from the works of \cite{kato2014,piersanti2014,wu2017,kato2018}; and \cite{wang2018}, which assume solar metallicity composition material in their nova calculations. At sub-solar metallicities the critical ignition mass increases slightly (acting to decrease the nova rate), the accretion efficiency increases substantially (acting to increase the WD growth rate and thereby increase the nova rate) and the nova regime shrinks (decreasing nova rates) \citep{chen2019}. With these competing influences, we assume that our current work probably only slightly over-predicts nova production as a result of neglecting the metallicity dependence of nova micro-physics. However, the systems which are most affected by this are low-metallicity, rapidly accreting nova systems. The vast majority of these nova systems produce prompt delay-time novae, meaning that the novae occur relatively soon after star formation. As low-metallicity star formation occurred in our Galaxy's distant past, most of these nova systems are unlikely to be observed in the modern Galaxy\footnote{Note that this is not necessarily true of stellar environments such as the Small and Large Magellanic Clouds, where star formation continues to occur at low metallicity.} anyway, and so we expect our predictions of current nova rates and distributions to be largely unaffected by including metallicity dependent accretion regimes, critical ignition masses, and accretion efficiencies.

In our model, we do not consider the observability of our novae, instead simply tracking whether or not the eruption occurred, and under what conditions. \cite{kato2013} find that at very low metallicities the evolution of nova eruptions can change drastically. Under the optically thick wind theory of nova outbursts \citep{kato1994}, which has proven very successful in modelling nova light curves \citep[e.g.,][]{hachisu2006}, novae rely on the Fe opacity peak to drive the nova winds. This means that novae at low metallicities tend have systematically slower light-curve evolution than their solar metallicity counterparts \citep{kato1997}. \cite{kato2013} find that at \Z=\tento{-3} the iron opacity peak declines such that it is roughly equal in magnitude to the lower temperature He opacity peak. Below this threshold the winds become far weaker. Below \Z=\timestento{4}{-4}, novae on WDs less massive than 0.7 M\solar\ are found to have negligible optically thick winds and have outbursts so slow that the extended stage of the outburst could last so long that they may be mis-classified as F-supergiants \citep{kato2013}. For WDs more massive than 0.7 M\solar, there are weak winds driven by the He opacity peak. These metallicity-dependent observability concerns have particular relevance in the context of globular clusters with \Z$\lesssim$\tento{-3} and other environments where the nova rates are expected to be dominated by low metallicity systems.

\subsection{The relative importance of metallicity-dependent environment modelling}

Most stellar environments where we might be interested in novae include multiple stellar populations, whether in the relatively discrete metallicity clumps such as those found in most globular clusters or the more continuous variation found in larger galaxies. In Section~\ref{sec:results,galaxies}, we present the results of our simulations convolved with age-metallicity-SFR models for the MW and M31. Further, in Figure~\ref{fig:eventratehistorym31} we explicitly address the effect of employing progressively simpler nova grids on nova event rate histories in M31. It is of interest to expand this discussion into a comparison between the relative importance of uncertainties in binary stellar evolution and building a detailed, well resolved, metallicity-dependent model of a star forming environment.

Producing and reducing data for the full 16-point metallicity grid used for this work costs around 360 CPU hours. Although this number may not appear excessive by the computational standards of many fields of astronomy, this data is only truly valid for a single case of binary physics. Ultimately, the goal would be to compute the results of many such physics cases and compare with observations of real stellar environments, but needing to compute each physics case for a range of metallicities increases the cost and difficulty of such a study significantly. 

Figure~\ref{fig:eventratehistorym31} demonstrates that a smaller metallicity grid of 7 points can achieve a satisfactorily similar event-rate history to our more expensive 16-point metallicity grid. However, if we were only interested in comparing with direct nova observations, then we might only care about the current nova properties of a given environment. There is relatively little variation in M31's nova rate predictions when varying the metallicity-grid used in the convolution, with rates varying only between 33-39 events per year. 

Roughly half the current nova rate in M31 can be attributed to systems with \Z>0.01, despite our adopted SFR model for M31 having a relatively high ratio between ancient and recent star formation. The fraction of novae originating from systems with \Z>0.01 increases to approximately 70 per cent in the MW. In this metallicity regime, the changes to the shape and magnitude of the delay-time distributions are small. This means that if a stellar environment is undergoing even relatively low levels of star formation, relying on the results of even a single metallicity -- close to the current metallicity of the environment -- can actually provide a good approximation for the evolutionary history of at least half of the currently observed nova systems. At the very least, uncertainties in the predicted nova rate caused by poorly resolving the metallicity evolution of the environment are unlikely to exceed 10 per cent. The inclusion of metallicity-dependent micro-physics is unlikely to change this, for the reasons discussed in Section~\ref{sec:chen2019disc}. 

In contrast, K21 found that varying the value of common envelope efficiency parameters resulted in the predicted nova rate in M31 to vary by up to a factor of 4. In this work, we have assumed the binary fraction to be 50 per cent, but in nature this fraction may be larger in the higher mass regions of our parameter space. Increasing the binary fraction to 100 per cent results in predictions for the current nova rates in the MW and M31 to increase to 60 and 66 novae per year respectively. A comprehensive investigation into the binary parameter space is yet to be conducted, but at this point it seems clear that, at least in terms of predicting event rates in spiral galaxies, uncertainties in modelling the metallicity-dependence of novae can be considered of secondary importance compared to uncertainties in binary stellar evolution.

In stellar environments where star-formation is neither continuous nor ongoing, such as elliptical galaxies and globular clusters, uncertainties in the metallicity evolution should be even less troublesome. The bursty nature of these stellar environments means that, provided the masses, ages, and metallicities of each burst are known, an extremely accurate approximation to the age-metallicity-SFR relation should be obtainable with relatively little effort.

However, in active star forming environments where relatively low metallicity star formation is ongoing, such as the Small and Large Magellanic Clouds, resolving the the metallicity-evolution should be expected to be of greater importance. At lower metallicities (\Z<0.01) the changes to the delay-time distributions are more pronounced, and so simple `current metallicity' approximations may not be a good model for the current state of nova populations.

All of the above discussion is relevant only for predictions of `current' nova properties as they may appear today. Central to the good agreement in current event rates between the different metallicity grids in Figure~\ref{fig:eventratehistorym31} is the fact that most of the novae from old stellar environments occurred in the distant past, and so are not observable. However, the cumulative effect of novae is of interest in the context of galactic chemical evolution. Here, the total nova productivity\footnote{It is in fact the total mass of nova ejecta -- or, even better, the total mass of nuclear-processed nova ejecta -- that is the most relevant quantity here.} of each star burst suddenly becomes relevant. Figure~\ref{fig:zsumtab} and Table~\ref{tab:zsumtab} summarise the relationship between the total nova productivity at each metallicity, showing that the nova productivity of a very low (e.g., \zml) metallicity star forming environment can be double that of a solar metallicity environment. We therefore expect that accurately modelling the metallicity dependence of novae to be far more important when considering nova yields.

In summary, we find that the impact of metallicity variation within a given stellar environment can be considered to be a minor source of uncertainty when making predictions of nova properties as they would appear today. A comprehensive assessment of the degree of variation caused by changing evolutionary parameters is needed before a more precise statement on their relative importance can be made. However, when considering the effects of novae on the chemical evolution of our Universe, the cumulative nature of this quantity leads us to the expectation that accurately treating the metallicity dependence of novae will be far more important.

\section{Conclusion}
\label{sec:conclusion}

In this work, we use the population synthesis code \binaryc\ to investigate the impact that metallicity-dependent stellar evolution has on nova rates and distributions. The predicted nova rates and distributions in this paper are for theoretical, intrinsic novae. Factors affecting the observability of these novae, such as the systematically slower evolution of nova light curves at low metallicities \citep{kato2013}, are not accounted for in our model. We present distributions of nova white dwarf masses, accretion rates, delay-times, and initial system properties for the two extremes of our 16-point metallicity grid, \zml\ and \zmu, in addition to describing the more nuanced morphological transitions within the grid. Using our full 16-point metallicity grid, we predict current nova rates and distributions for the Milky Way and M31. We discuss our findings in the context of uncertainties in binary stellar evolution, metallicity variation inherent to nova micro-physics, and observed nova rates and distributions.

We find a clear anti-correlation between the total nova productivity of a burst of star formation and metallicity. Nova productivity in the highest-metallicity star bursts (\zmu) is roughly half that of the lowest (\zml). This trend is primarily driven by metallicity-associated reductions in stellar radii causing increased contributions from novae originating from low-mass white dwarfs with giant donors. A notable secondary influence is the effect of the systematically higher remnant masses at lower metallicities. This allows wind-accreting systems in particular to occur in new regions of the parameter space\footnote{Here we refer to the parameter space of initial conditions, specifically the initial primary mass, secondary mass, and the orbital separation.} at lower metallicities, in addition to increasing nova productivity in regions of shared parameter space due to reducing critical ignition masses.

Metallicity has a clear influence on the distributions of white dwarf masses, but only very minor effects on accretion rate distributions. At lower metallicities, 0.6-1 M\solar\ C/O white dwarf contributions increase, with most of these new novae being produced in the presence of a giant donor. We find that first (red) giant branch (FGB) and thermally pulsing asymptotic giant branch (TPAGB) donor stars remain dominant across all metallicities, contributing over half of all nova events. The relative importance of FGB and TPAGB donor stars increases slightly with metallicity.

As metallicity reduces, stellar life times tend to be shorter, the effect of which can be most clearly observed in our nova delay-time distributions. The delay-time distributions themselves vary relatively little in shape with metallicity, the main difference being that lower metallicities have systematically stronger nova production at delay-times between 1-3 Gyr. This change is driven by the more prompt introduction of FGB-donor systems, which occurs approximately 1.5 Gyr sooner at \zml\ compared to \zmu.

Comparing low- and high-metallicity delay-time distributions also reveals that there are systematic differences in the distributions of white dwarf compositions. At \zml, C/O white dwarf contributions make up 40 per cent of all novae and 80 per cent of novae with delay-times longer than 10 Gyr, while at \zmu, C/O white dwarf contributions represent only 25 per cent of all novae, and only 40 per cent of novae with delay-times longer than 10 Gyr. 

We compute a current Galactic nova rate of approximately 33 novae per year, broadly consistent with observational estimates making use of extra-Galactic nova data. This prediction is inconsistent with observational estimates making use of Galactic nova data, including the recent \cite{de2021} estimate of $43.7^{+19.5}_{-8.7}$ novae per year, which used Galactic novae observed in the near-infrared. However, we note that our theoretical rate estimate could potentially vary by as much as 30 novae per year under different combinations of physical assumptions.

We find that 70 per cent of the currently observable MW novae occur in the thin disk, 24 per cent in the bulge, 5 per cent in the thick disk, and 1 per cent in the halo. For each Galactic component we present predicted distributions for accretor-donor stellar types, white dwarf masses, accretion rates, and metallicity. We find that the presence -- or absence -- of recent/current star formation dramatically affects the distributions of the nova white dwarf masses and metallicities in particular. The observed sensitivity of these distributions to the evolution histories of the different components makes novae potentially useful tools to probe the evolution of distant stellar environments.

In M31, we compute a current nova rate of approximately 36 novae per year, lower than $65^{+15}_{-16}$ novae per year \citep{darnley2006}, the most recent observational estimate of M31's nova rate. We find that, provided only estimates of currently observable nova properties are of interest, our 16-point metallicity grid is excessive for most stellar environments. For stellar environments where relatively high metallicity (\Z>0.01) star formation is ongoing, or where star formation has ceased in the distant past, selecting a single metallicity close to the current metallicity of the environment should provide an acceptable approximation of the currently observable nova rates and distributions. Regarding non-current nova properties, we find that we are able to approximate the event-rate history of M31 very well with a grid of 7 carefully selected metallicities.

In the context of the large parameter space and associated uncertainties surrounding binary stellar evolution \citep{demarco2017}, we conclude that, despite the presence of clear relationships between nova productivity and the metallicity of the star burst, the effect of metallicity on novae is likely of only secondary importance. However, this result is unlikely to hold in stellar environments undergoing active star formation at low metallicity, such as the Small and Large Magellanic Clouds. Further, if quantities that are dependent on cumulative nova production are of interest, such as the chemical yields from nova outbursts, then a good treatment of the metallicity evolution of the stellar environment is also necessary.

\section*{Acknowledgements} 

The authors wish to thank M. Kato and B. Wang for the provision of data used in this work.
A.~R.~C. is supported in part by the Australian Research Council through a Discovery Early Career Researcher Award (DE190100656). CK acknowledges funding from the UK Science and Technology Facility Council (STFC) through grants
ST/R000905/1 and ST/V000632/1.
R. ~G. ~I. thanks the STFC for funding, in particular Rutherford fellowship ST/L003910/1 and consolidated grant ST/R000603/1.
Parts of this research were supported by the Australian Research Council Centre of Excellence for All Sky Astrophysics in 3 Dimensions (ASTRO 3D), through project number CE170100013.
This research was supported in part by a Monash University Network of Excellence grant (NOE170024).

\section*{Data Availability}

The data underlying this article will be shared on reasonable request to the corresponding author.



\bibliographystyle{mnras}
\bibliography{INbib.bib} 







\bsp	
\label{lastpage}
\end{document}